\documentclass[fleqn,usenatbib]{mnras}

\usepackage{newtxtext,newtxmath}

\usepackage[T1]{fontenc}

\DeclareRobustCommand{\VAN}[3]{#2}
\let\VANthebibliography\thebibliography
\def\thebibliography{\DeclareRobustCommand{\VAN}[3]{##3}\VANthebibliography}

\usepackage{graphicx}	
\usepackage{amsmath}	

\title[Gravitational wave spectral synthesis]{Gravitational wave spectral synthesis}

\author[W. G. J. van Zeist et al.]{
Wouter G. J. van Zeist,$^{1,2,3}$\thanks{E-mail: wouter.vanzeist@astro.ru.nl} J. J. Eldridge$^{3}$ and Petra N. Tang$^{3}$
\\
$^{1}$Department of Astrophysics/IMAPP, Radboud University, PO Box 9010, 6500 GL, Nijmegen, The Netherlands\\
$^{2}$Leiden Observatory, Leiden University, PO Box 9513, 2300 RA, Leiden, The Netherlands\\
$^{3}$Department of Physics, University of Auckland, Private Bag 92019, Auckland, New Zealand
}

\date{Accepted XXX. Received YYY; in original form ZZZ}

\pubyear{2023}

\begin{document}
\label{firstpage}
\pagerange{\pageref{firstpage}--\pageref{lastpage}}
\maketitle

\begin{abstract}
We study the LISA sources that arise from isolated binary evolution, and how these depend on age and metallicity, using model stellar populations from \textsc{bpass}. We model these as single-aged populations which are analogous to star clusters. We calculate the combined GW spectrum of all the binaries within these model clusters, including all types of compact binaries as well as those with living stars. These results allow us to evaluate the detectability of star clusters with LISA. We find at late times the dominant sources are WD–WD binaries by factors of 50–200, but at times between $10^8$ and $10^9$ years we find a significant population of NS–WD and BH–WD binaries (2–40 per $10^6$ M$_{\odot}$), which is related to the treatment of mass transfer and common envelope events in \textsc{bpass}, wherein mass transfer is relatively likely to be stable. Metallicity also has an effect on the GW spectrum and on the relative dominance of different types of binaries. Using the information about known star clusters will aid the identification of sky locations where one could expect LISA to find GW sources.
\end{abstract}




\section{Introduction}

Ever since the first direct detection of gravitational waves (GWs) in 2015 \citep{gw150914}, the observation of GWs by the LIGO–Virgo–KAGRA (LVK) Collaboration has proven a highly fruitful venture. As of the end of its third observing run in 2020, this collaboration has detected the 90 GW transients \citep{gwtc3}, each originating from a binary merger of black holes (BHs) or neutron stars (NSs), including multiple individually notable events such as GW170817 \citep{gw170817_1}, a binary NS merger for which an associated gamma-ray burst and optical transient were also detected \citep{gw170817_2}, opening up a new field of astronomical research known as multi-messenger astronomy.

However, the currently operating GW observations have a limitation in that they cannot detect GWs at frequencies below about 10 Hz, which is in part because they are located in contact with the surface of the Earth, and thus also pick up seismic vibrations, which contribute noise to the detectors at low frequencies \citep{saulson1984,hughes1998}, and because the intrinsic sensitivity of an interferometric GW detector drops when the wavelength becomes much longer than its arms. As a result, there are many types of potential GW sources that the detectors of the LVK Collaboration could not observe, such as those from white dwarf (WD) binaries or from mergers of (super)massive BHs.

To survey events below the ground-based detectors' frequency range, a space-based GW observatory called the Laser Interferometer Space Antenna (LISA) is being developed, expected to be launched in the 2030s. LISA will consist of three satellites in a triangular formation with a separation of 2.5 million km, with the system as a whole following Earth's orbit but trailing by approximately 20 degrees \citep{lisa_l3,cornish2017}. The three satellites will comprise a system of three overlapping double-arm interferometers.

The frequency range in which LISA will be able to detect GWs spans roughly from $10^{-4}$ to 1 Hz. Within this range, aside from the aforementioned compact object binaries and massive BH mergers, there are also numerous other potential GW sources that have been theorised, such as extreme mass-ratio inspirals (EMRIs), which involve a stellar-mass object falling into a massive BH \citep{amaro2007}; or systems of a living star (i.e. a star with nuclear fusion as the main source of energy production) being envelope-stripped by a WD or NS, which could also be X-ray sources \citep{yungelson2008,goetberg2020}. A detailed analysis of astrophysical sources potentially visible to LISA can be found in \citet{lisa_astrophysics}; there are also potential cosmological sources, which are discussed in \citet{lisa_cosmology}.

Apart from LISA, there are several other planned space-based GW observatories which would also survey the milli-Hz range, including TianQin \citep{tianqin} and Taiji \citep{taiji}. One proposed detector in particular that we will look at is $\mu$Ares, which would use a similar design and technology to LISA, but with arm lengths of 100 million km instead, lying within the orbit of Mars \citep{muares}. This detector, which is in the early stages of design and may be launched in the decades following LISA, would have a frequency range overlapping that of LISA but extending down to $10^{-6}$ Hz, which would include wider compact object binaries but also particularly more living–compact binaries, which tend to lie at the lower end of LISA's frequency range due to physical limits on their separation.

Understanding the signals that would be received by LISA necessitates modelling the sources that could contribute to the LISA GW signal. There have been many analyses that model the compact binary population within the Milky Way \citep[e.g.][]{portegies1998,nelemans2001,lamberts2018,lamberts2019,breivik2020}. However, within this population there will be many stellar clusters containing binaries that can be detected by LISA. Previous studies have considered binaries formed by dynamical interactions \citep[e.g.][]{gcbhejection2,lisa_gc1,ivanova2006,lisa_gc2,lisa_gc3,lisa_gc4,lisa_gc5}, however not all clusters will have high densities that drive dynamical interactions. So a study of sources that arise from isolated interacting binary evolution is important to understand the majority of LISA sources.

In particular, we make a distinction between two types of stellar clusters, open and globular clusters. Open clusters have a relatively low mass and density, which means that separate stellar systems in the cluster will not interact with each other frequently, and thus can be modelled quite closely by isolated binary evolution. Globular clusters have more mass and thus more potential GW sources, but because of their high density there will be dynamical interactions between systems, and thus the assumption of isolated binary evolution is less accurate. An overview of predictions of LISA sources in both types of clusters can be found in section 1.3.2 of \citet{lisa_astrophysics} and references therein.

In this paper, we study the LISA sources that arise from isolated binary evolution, and how these depend on the age and metallicity of the simple stellar population. We include all combinations of BHs, NSs and WDs, but also living binaries and living–compact binaries like those of \citet{goetberg2020}. To do this, we use model stellar populations of various ages and metallicities from the population synthesis code suite \textsc{bpass} \citep[Binary Population and Spectral Synthesis,][]{bpass1,bpass2}, and calculate the combined GW spectra of all the binaries within these populations. We focus on sources in the LISA frequency range, with some discussion of $\mu$Ares.

Our work is analogous to the many studies that predict electromagnetic (EM) spectra for unresolved simple stellar populations. These have a long history of being used to understand these stellar populations \citep[e.g.][]{tinsley1968,wofford2016,bpass2}.

The \textsc{bpass} stellar populations we are using consist of systems formed in a single starburst and therein resemble a stellar cluster where dynamical interactions are rare. We use our models to investigate whether a stellar population similar in mass and distance to a globular cluster would be detectable by LISA and what the dominant sources in the stellar population would be.

The outline of this paper is as follows: in section \ref{chapter_method} we outline our method of simulating GWs and producing the model populations; in section \ref{chapter_results} we present our GW population synthesis and the results in terms of frequency spectra, importance of the different types of binaries, the effect of age and metallicity on the stellar population, and how realistic these predictions may be; in section \ref{chapter_examples} we illustrate our results with example calculations for real clusters; and in section \ref{chapter_discussion} we discuss the results and present our conclusions.

\section{Methods of Gravitational Waveform Synthesis} \label{chapter_method}

\subsection{Simulating gravitational waves}

Identifying GW signals in data requires template models of what these waveforms would be expected to look like. The mathematical theory that is used in simulations of GWs is based upon the Einstein field equations, which are difficult to solve both analytically and numerically. Consequently, models and simulations of GWs use approximations and numerical methods to simplify the problem and reduce the difficulty of solving the Einstein equations while retaining as much accuracy as needed.

When simulating the gravitational waveforms from mergers of black holes or neutron stars that would be detectable by LIGO, particular methods of simplification that are commonly used include post-Newtonian approximations and limited numerical relativity. Such methods are used, for example, in LIGO's code suite \textsc{LALSuite} \citep{lalsuite}, our own GW simulation package \textsc{Riroriro} \citep{riroriro,bpassmassdist} and the LISA-targeted package \textsc{legwork} \citep{legworkjoss,legworksci}.

However, binaries in the frequency range of LISA have significantly wider orbits than those detectable by LIGO, so a simpler waveform model can be used, as the periods of these binaries effectively do not evolve over the time of observation. Furthermore, because we do not need to simulate the process of the binaries merging there are fewer relativistic effects that need to be taken into account.

Our method we use to create our simulations of LISA-detectable binaries is as follows. Firstly, the (dimensionless) instantaneous amplitude of the GW signal of a binary in the frequency range of LISA can be written as a single equation \citep{krolak2004,shah2012}:
\begin{equation} \label{instamp}
	\mathcal{A} = \frac{4 (G M_{\rm chirp})^{\frac{5}{3}}}{c^{4} d} (\pi f)^{\frac{2}{3}}
\end{equation}

In this equation, $G$ is the gravitational constant, $c$ is the speed of light, $d$ is the distance between the binary and the observer, $f$ is the gravitational wave frequency of the binary (equal to twice the orbital frequency) and $M_{\rm chirp}$ is the chirp mass of the binary:
\begin{equation}
	M_{\rm chirp} \equiv \frac{(m_{1} m_{2})^{\frac{3}{5}}}{(m_{1} + m_{2})^{\frac{1}{5}}}
\end{equation}

From this, we can compute the binary's characteristic strain, a simple way to evaluate the strain accumulated by a binary as it is observed over time \citep{moore2015,kupfer2018}:
\begin{equation} \label{charstrain}
	h_{c} = \sqrt{f T_{\rm obs}} \mathcal{A}
\end{equation}

Here, $T_{\rm obs}$ is the length of time for which the binary is observed. Another common parameter used to evaluate a GW signal observed over time is the strain power spectral density (PSD), which is as follows \citep{moore2015}:
\begin{equation} \label{psd}
    S_{h} = \frac{h_{c}^2}{f} = T_{\rm obs} \mathcal{A}^2
\end{equation}

The characteristic strain and the PSD both have properties which make them advantageous as indicators of the ``loudness'' of a GW signal in a detector: a source's characteristic strain over frequency can be plotted against a detector's sensitivity curve, and the height between these curves is then directly related to the SNR of that GW in that detector; while a source's PSD, integrated over frequency, gives the mean square amplitude in the detector \citep{moore2015}. The characteristic strain is dimensionless, while the PSD has units of $f^{-1}$.

In this paper, we will use the PSD as the primary method to quantify strain in our plots, but we will also show the equivalent plots with characteristic strain in the appendix.

The calculation of instantaneous amplitude in equation \ref{instamp} assumes that the binary has an optimal inclination angle; to account for arbitrary inclinations, the amplitude can be averaged over all possible inclinations, giving the following multiplication factor for $\mathcal{A}^2$ \citep{robson2019}:
\begin{equation}
    \frac{1}{2} \int_{-1}^{1} \left(\frac{(1 + x^2)^2}{4} + x^2 \right) dx = \frac{4}{5}
\end{equation}

Therefore, when we calculate the strain of a binary using equation \ref{charstrain} or \ref{psd} we multiply $\mathcal{A}$ by $\sqrt{4/5}$.

We treat all binaries as having zero eccentricity when estimating their strain, using the assumption that these binaries would likely circularise quickly during their evolution. For systems were the second compact object formed is a WD, there is no SN and therefore the eccentricity will be low. For systems where the second compact object is a BH or NS, for systems in the LISA frequency range the periods would be short enough that the SN must have had a low ejecta mass and the object would have received a low natal kick, and therefore we assume that the orbit would not have been significantly affected and the eccentricity would be low \citep{bpasskick}.

We note that \citet{eccentricity_lau} and \citet{eccentricity_wagg} show that some NS and BH binaries may retain some non-negligible eccentricity. However, highly eccentric systems can merge much quicker than those with circular orbits and the emission of GWs typically circularizes the orbit faster than the period of the orbit decreases \citep{peters1964}. However, non-negligible eccentricities, $e \sim 0.1 \times {\rm few}$, may remain and would alter the expected signal from the binaries. We do not consider these in our work and this should be reviewed in future.

\subsection{Sensitivity of LISA}

To evaluate the detectability of a GW signal with LISA, one needs to compare it to LISA's sensitivity, which is limited by instrument noise, confusion noise and its anisotropic detector response function. Specifically, a prediction of the GW sensitivity of LISA with respect to frequency, in units of PSD (Hz$^{-1}$) is given by the following equation \citep{robson2019}:
\begin{equation} \label{sc_instrument}
    S_n(f) = \frac{10}{3L^2} \left(P_{\rm OMS} + \left(1 + \cos^2\left(\frac{f}{f_*}\right)\right) \frac{2 P_{\rm acc}}{\left(2 \pi f\right)^4}\right) \left(1 + \frac{6}{10} \left(\frac{f}{f_*}\right)^2\right)
\end{equation}

In this equation, $L$ is the arm length of LISA (2.5 Gm), $f_*$ is a quantity called the transfer frequency which is defined by $f_* = c/(2\pi L)$ and is equal to 19.09 mHz for LISA, $P_{\rm OMS}$ is the (high-frequency) single-link optical metrology noise \citep{robson2019,lisanoisedocument}:
\begin{equation}
    P_{\rm OMS} = \left(3 \times 10^{-22}\right) \left(1 + \left(\frac{2 \times 10^{-3}}{f}\right)^4\right)
\end{equation}

And $P_{\rm acc}$ is the (low-frequency) single-test-mass acceleration noise:
\begin{equation}
    P_{\rm acc} = \left(6 \times 10^{-30}\right) \left(1 + \left(\frac{4 \times 10^{-4}}{f}\right)^2\right) \left(1 + \left(\frac{f}{8 \times 10^{-3}}\right)^4\right)
\end{equation}

Both $P_{\rm OMS}$ and $P_{\rm acc}$ are given in units of Hz$^{-1}$.

Detector noise is not the only factor limiting the sensitivity of a GW detector, as this is also affected by the instrument not being equally sensitive in all directions. This anisotropy, referred to as the antenna beam pattern, is quantified by a detector response function, which also takes into account the detector's rotation over time.

While the beam-pattern equations for LISA \citep{cutler1998} are broadly similar to those used for ground-based detectors like LIGO \citep{projection1,projection2,projection3}, there are different conventions in how these are used. In LIGO literature, the detector response function is applied to the amplitude of the signal and not to the sensitivity curve, which contains only the detector noise; but in LISA literature this factor is applied to the sensitivity curve instead \citep{robson2019}.

Equation \ref{sc_instrument} includes a factor of the sky- and polarisation-averaged detector response, and so we do not need to take the detector response into account separately so long as we assume that each binary in our population has a random polarisation and the sky location of our population is also random. It is worth noting that, were one to look at an individual binary with a specific sky location and polarisation, then this averaged response would not apply; for information on how to calculate the detector response function of an individual source, see \citet{cutler1998,cornish2003,krolak2004}.

The sensitivity curve described by equation \ref{sc_instrument} solely includes noise sources related to the instrument itself, but LISA will also experience effective noise from unresolved galactic binary sources. This confusion noise, in units of PSD (Hz$^{-1}$) is described by the following equation \citep{cornish2017,robson2019}:
\begin{equation} \label{sc_confusion}
    S_c(f) = (9 \times 10^{45}) f^{-\frac{7}{3}} e^{-f^\alpha + \beta f \sin(\kappa f)} [1 + \tanh(\gamma(f_k - f))]
\end{equation}

In this equation, $\alpha,\beta,\kappa,\gamma,f_k$ are fit parameters which depend on the duration of the mission, given in table 1 of \citet{cornish2017}. A ``full'' sensitivity curve including both instrument and sensitivity noise is obtained by summing equations \ref{sc_instrument} and \ref{sc_confusion}.

\subsection{\textsc{bpass} stellar population models}

In order to simulate gravitational waves from stellar populations, we require data describing such a population to use as input. For this we use the Binary Population and Spectral Synthesis (\textsc{bpass}) code suite, which simulates the evolution of a population of binary and single-star systems from a wide range of initial conditions. We use v2.2.1 stellar models \citep{bpass1,bpass2} with a modified version of the gravitational wave population synthesis code \textsc{tui}, first discussed in \citet{bpassmassdist}. This is the computer program that calculates the birth and evolution (via gravitational radiation) of the population of gravitational wave sources and transients in the \textsc{bpass} code suite.

In creating the population we use the fiducial \textsc{bpass} parameters; that is, an initial mass function (IMF) based on \citet{kroupa1993}. This has a minimum mass of 0.1~M$_{\odot}$, with an IMF slope of -1.30 up to 0.5~M$_{\odot}$, where the slope steepens to -2.35 up to a maximum stellar mass of 300~$M_{\odot}$. The initial binary parameters are taken from \citet{moe2017} (see their table 13). This has a binary star fraction of 94 per cent above an initial primary mass of 16~M$_{\odot}$, which drops to 40 per cent for Solar-mass stars. The period and mass-ratio distributions for the orbits are observationally derived and too complex to summarise here \citep[for full details, see table 13 of][]{moe2017}. 

The binary stellar evolution is as described in \citet{bpass1} and \citet{bpass2}. All orbits during the stellar evolution of the binaries are here taken to be circular. This is a reasonable approximation as binaries with similar semilatera recta have similar evolutionary outcomes \citep{portegies1996,hurley2002}. The remnant masses are determined by the standard prescription from \citet{eldridge2004}. When a star undergoes a supernova we use a Maxwellian velocity distribution from \citet{hobbs2005}, which is adjusted to a momentum distribution for black holes, to simulate the resulting kick.

For the GW source calculations we search through all our stellar models at a certain metallicity, determining the chirp mass and orbital frequency for such systems and recording them, either when both stars are living, or when one has experienced a supernova and we now have a binary with a compact remnant. When both stars in a binary have formed a compact remnant we assume that only gravitational radiation drives the further evolution and integrate the equations of \citet{peters1964} to determine the chirp mass and frequencies of the double compact object evolution. If the orbit after the second supernova becomes eccentric we do model the evolution of eccentricity and period together.

The resulting outputs are a dataset of gravitational wave sources distributed in chirp mass, frequency and time since birth of the stars on the zero-age main sequence (ZAMS). These can then be combined with the above modelling of the gravitational wave signals to produce the expected results from a simple stellar population (i.e. one that contains only single stars and isolated binaries).

\section{Gravitational wave spectra of stellar populations} \label{chapter_results}

In this section we discuss the concept of ``gravitational wave spectral synthesis'', the simulation of gravitational waves from aggregated stellar populations generated by a population synthesis code, and show various results from these simulations for stellar populations with an initial mass of $10^{6} M_{\odot}$ formed in a single starburst, of various ages and metallicities, observed at 1 kpc for a duration of 4 years.

\subsection{Gravitational wave spectral synthesis}

The concept of gravitational wave spectral synthesis involves calculating the total frequency spectrum of the GWs of each binary (at a given age and metallicity) in a data set describing a stellar population, such as the previously discussed \textsc{bpass} model data set, added together. This gives the cumulative gravitational wave spectrum of a stellar population, analogous to the electromagnetic spectral synthesis already performed by \textsc{bpass}, wherein cumulative electromagnetic spectra of stellar populations are constructed. In both cases, looking at the spectra of stellar populations as a whole gives insights when it is not possible to resolve individual sources within them.

For each binary model in the data set from \textsc{bpass}, we calculated the GW strain using equations \ref{charstrain} and \ref{psd}. To create a cumulative GW spectrum from these individual strain values, we sorted the systems for each age and metallicity in the \textsc{bpass} data set into bins based on frequency. Within each bin, we summed the individual strains in quadrature (instead of linearly, which would only be valid if all of the binaries in the bin are in phase with each other). The resulting sum gives the strain amplitude for that bin. When we need to calculate the strain power, we square this value.

\subsection{Spectra of different types of compact binaries} \label{sec_compact}

\begin{figure*}
    \centering
    \includegraphics[width=1\columnwidth]{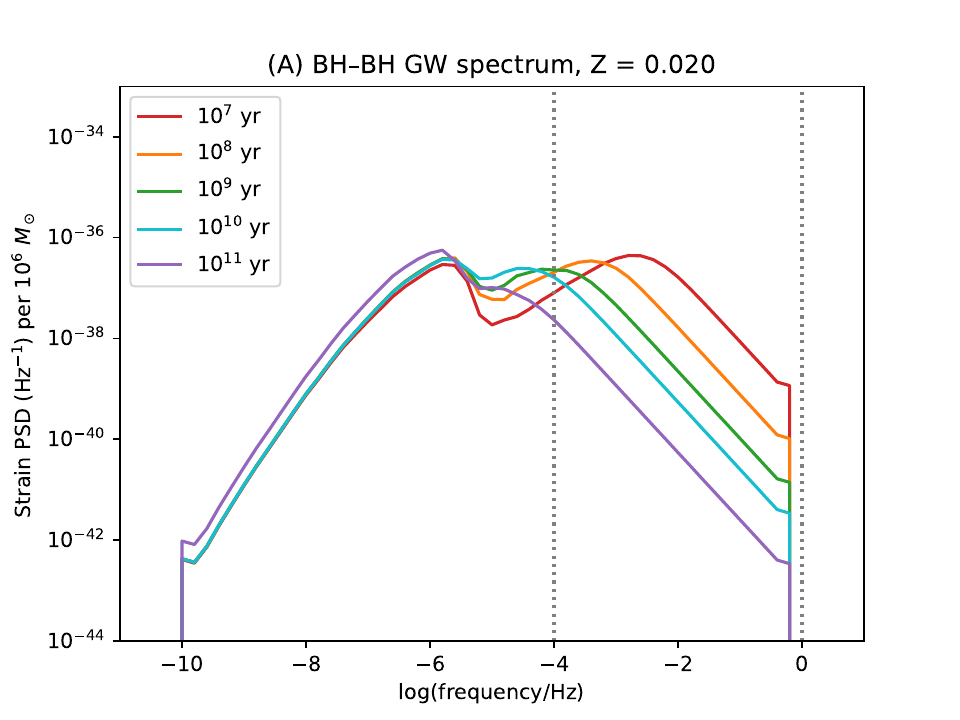}
    \includegraphics[width=1\columnwidth]{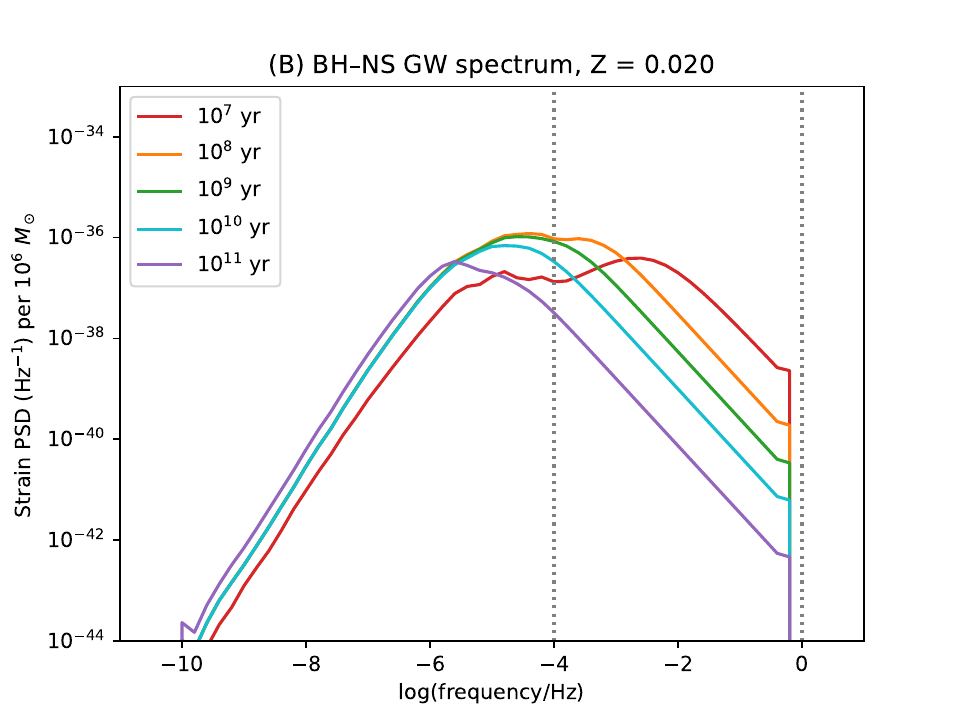}
    \includegraphics[width=1\columnwidth]{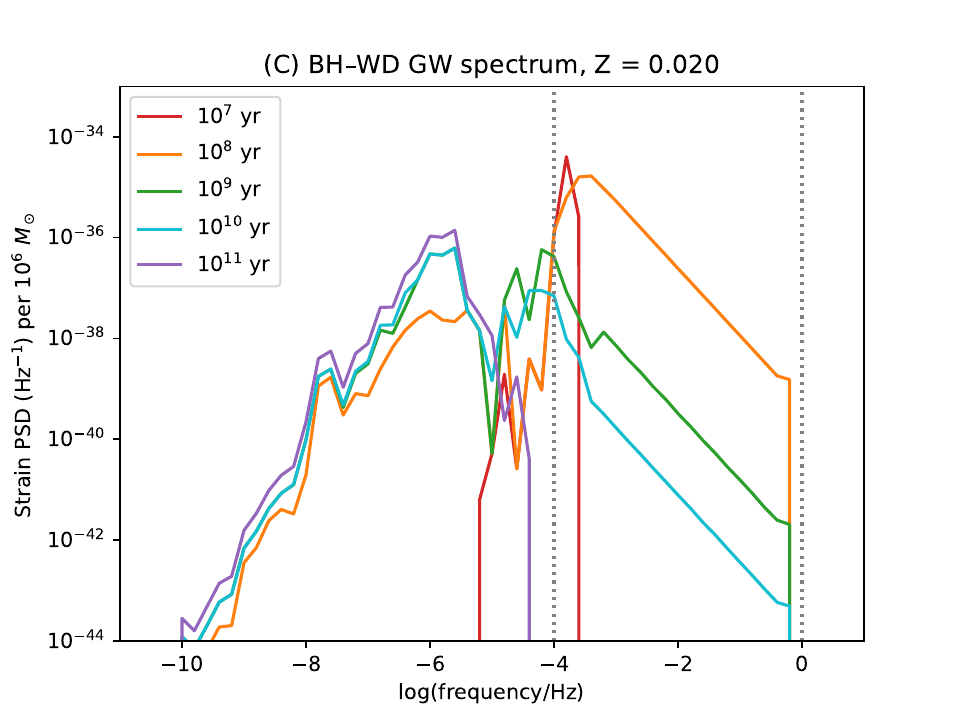}
    \includegraphics[width=1\columnwidth]{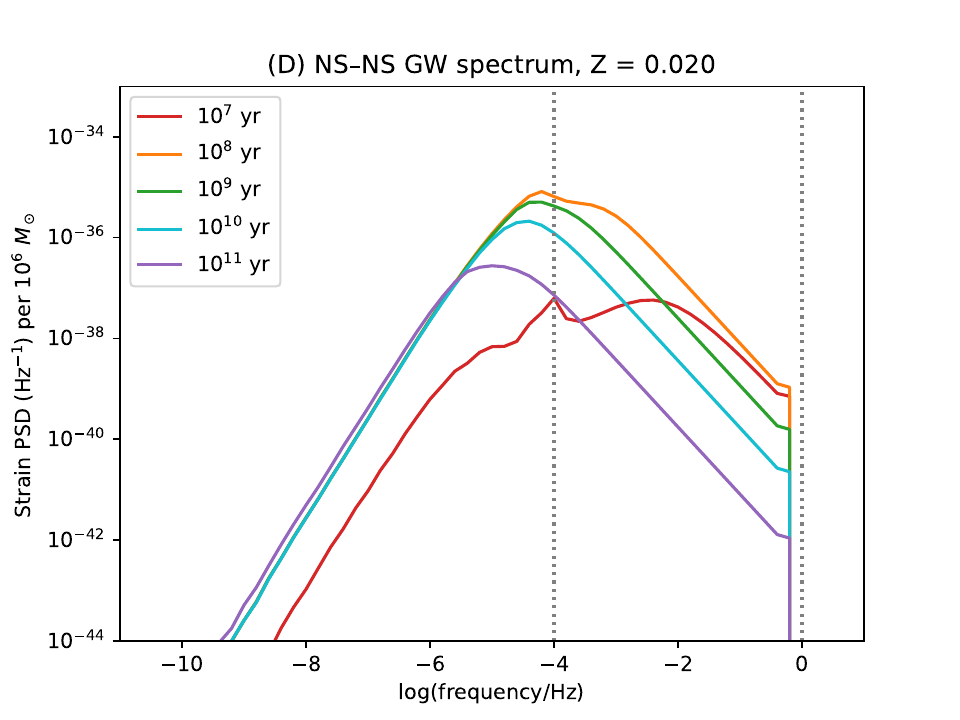}
    \includegraphics[width=1\columnwidth]{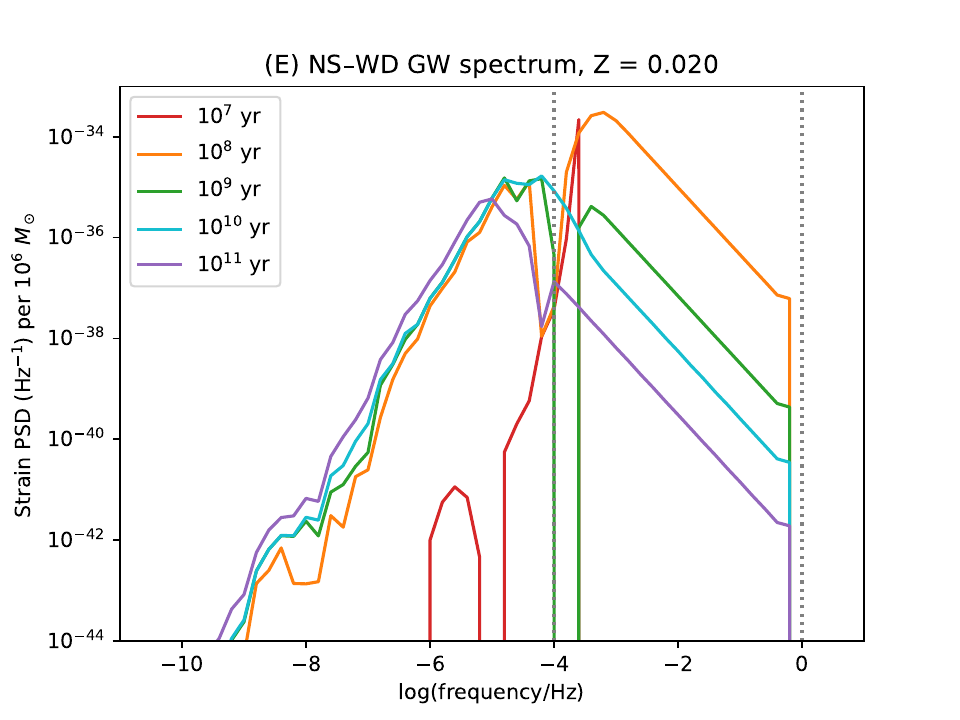}
    \includegraphics[width=1\columnwidth]{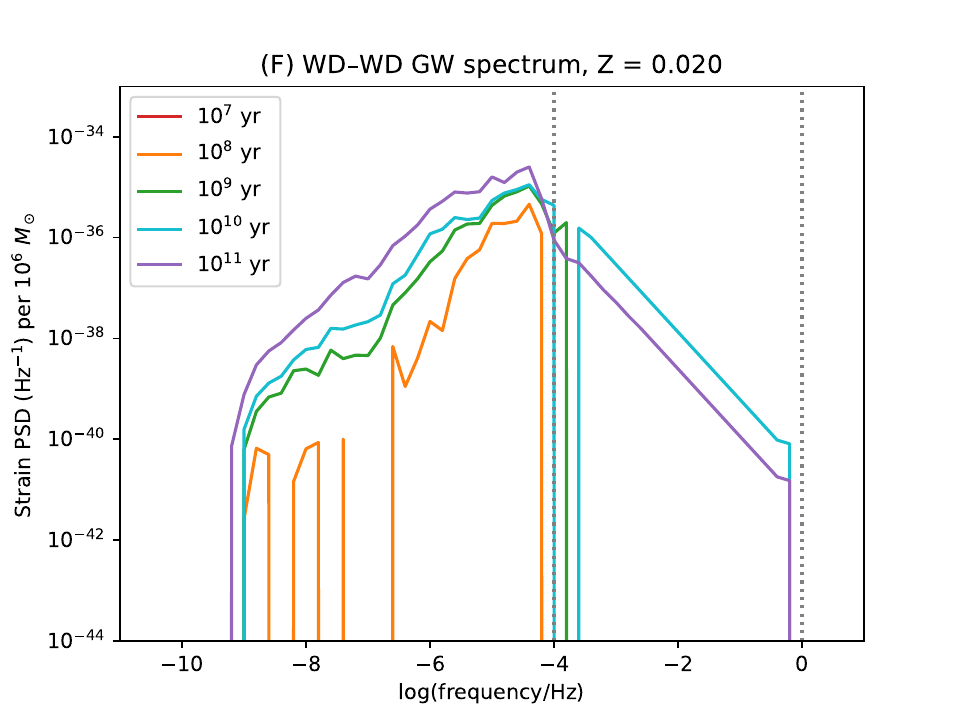}
    \caption{Gravitational wave spectra of different types of compact binaries for a stellar population with an initial mass of $10^6$ M$_{\odot}$ at a distance of 1 kpc and $Z$ = 0.020. Each panel shows a different binary type, and each line a different point in time after the initial starburst. The dotted lines show an approximation of the LISA frequency range.}
    \label{compact_types}
\end{figure*}

\begin{figure*}
    \centering
    \includegraphics[width=1\columnwidth]{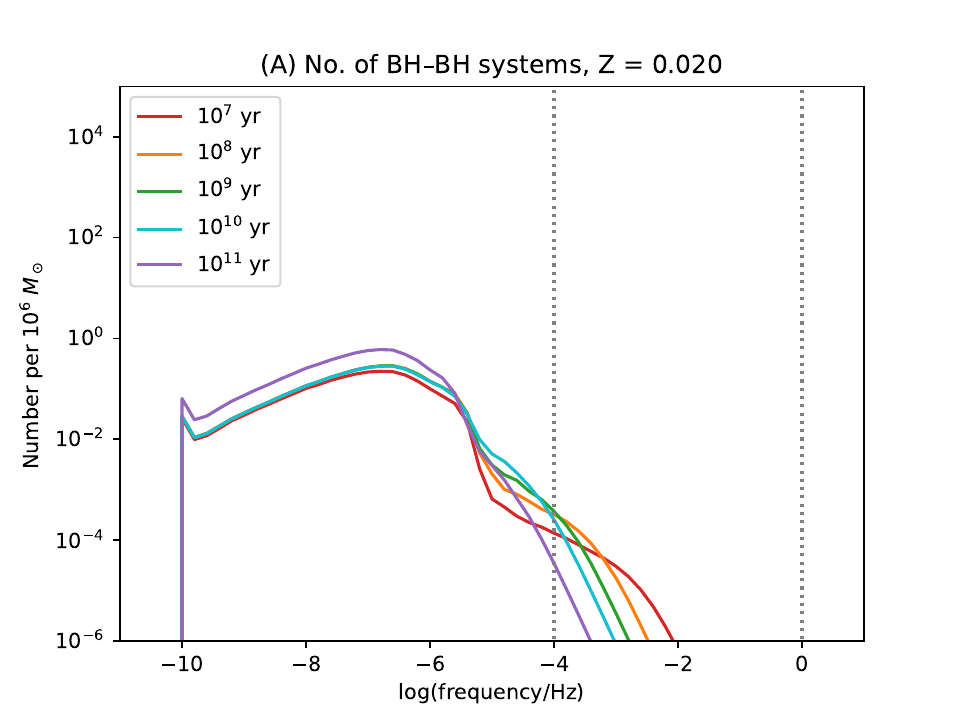}
    \includegraphics[width=1\columnwidth]{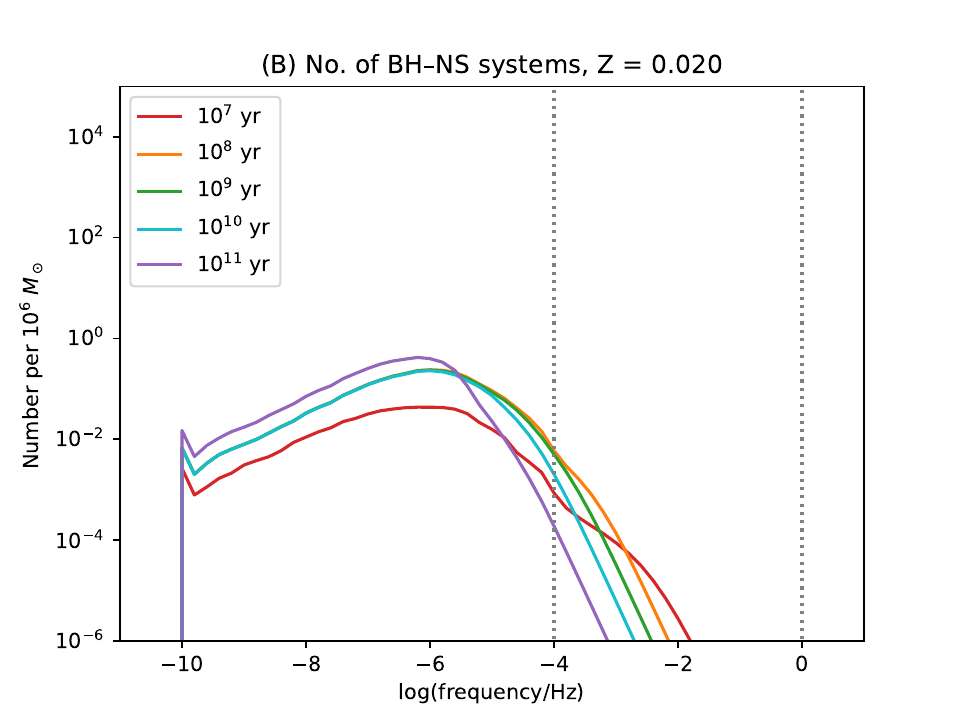}
    \includegraphics[width=1\columnwidth]{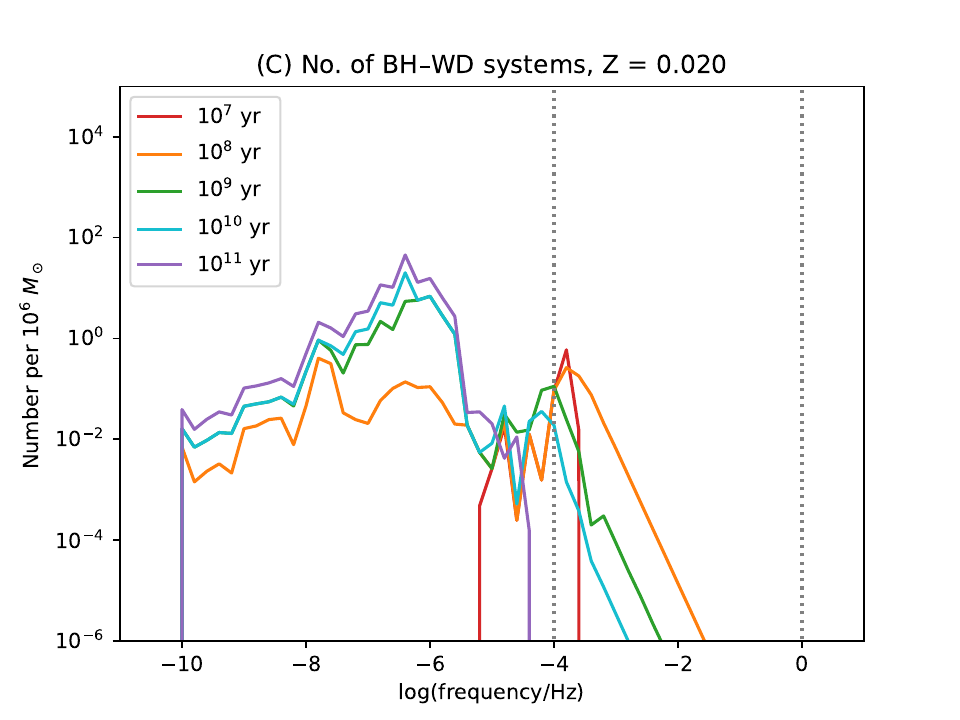}
    \includegraphics[width=1\columnwidth]{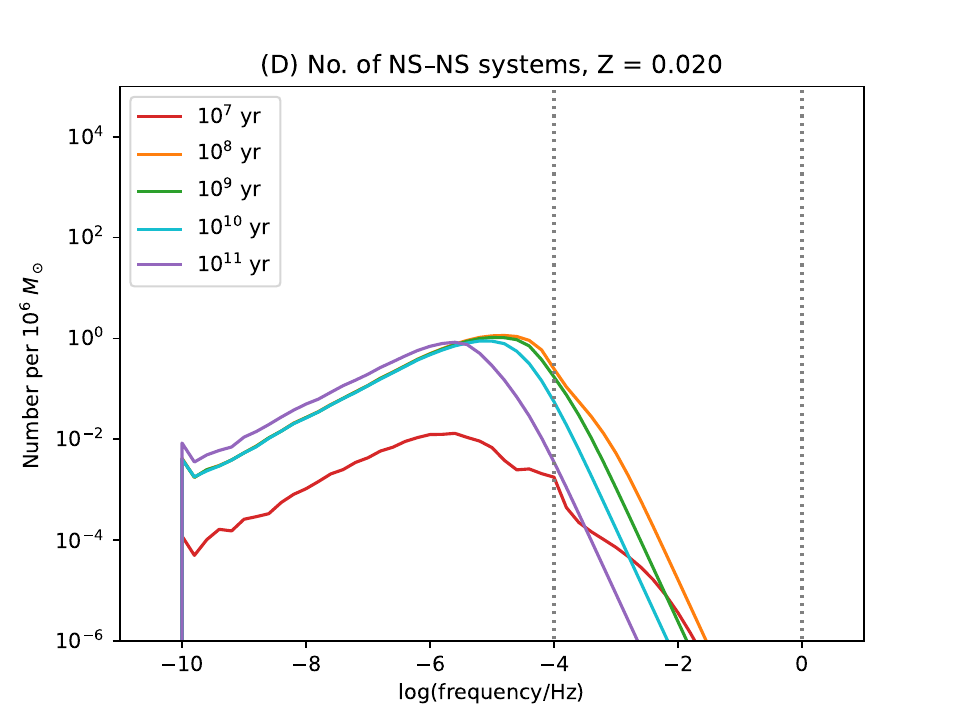}
    \includegraphics[width=1\columnwidth]{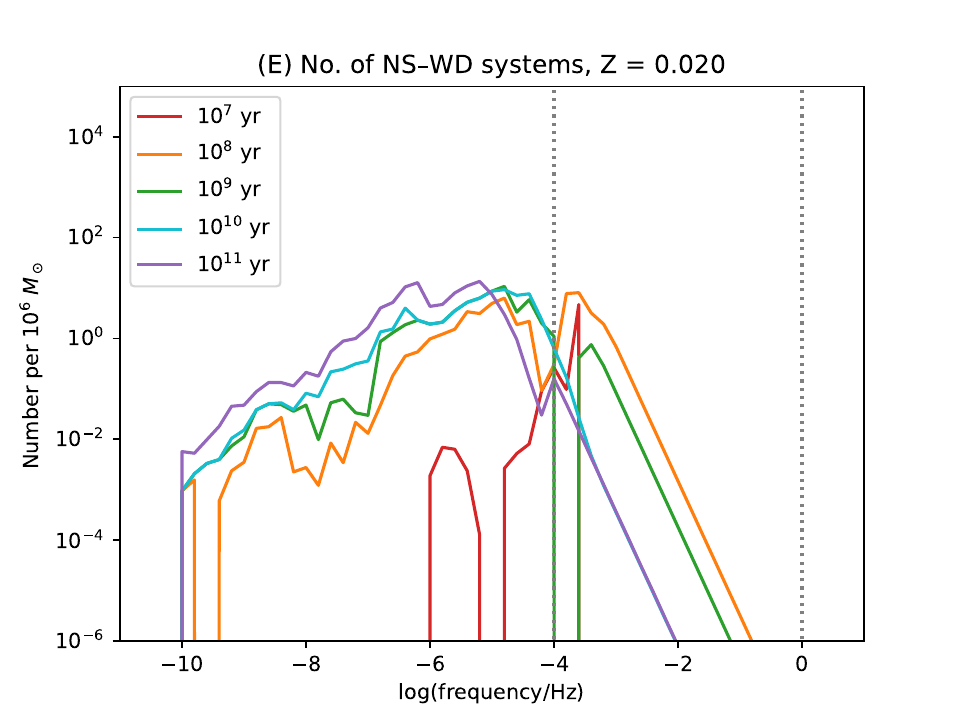}
    \includegraphics[width=1\columnwidth]{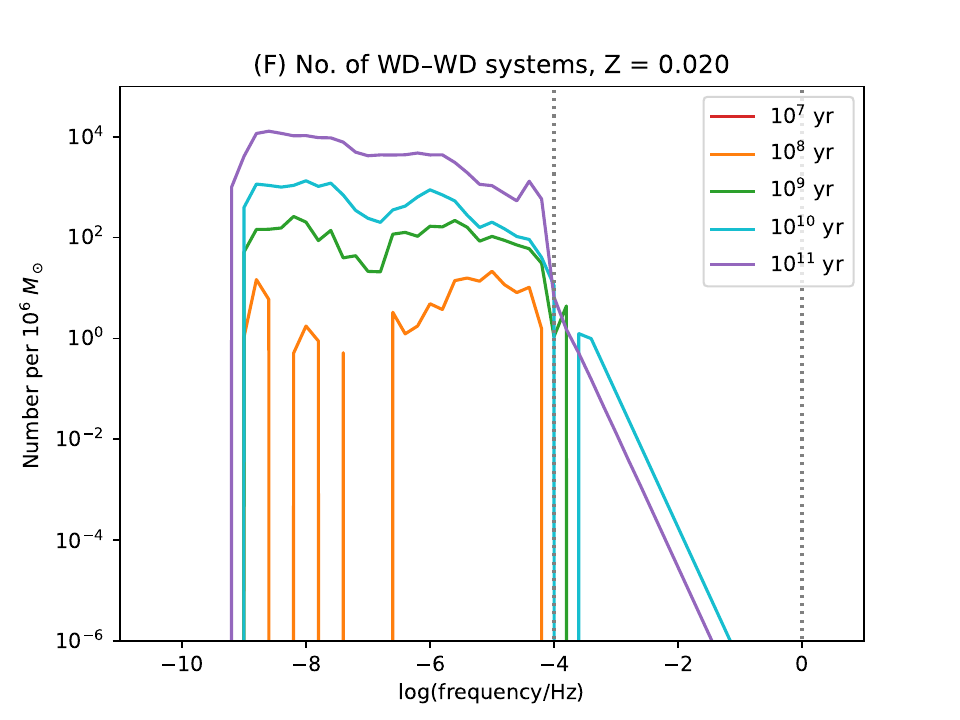}
    \caption{Number of systems per bin for different types of compact binaries for a stellar population with an initial mass of $10^6$ M$_{\odot}$ at $Z$ = 0.020. Each panel shows a different binary type, and each line a different point in time after the initial starburst. The dotted lines show an approximation of the LISA frequency range.}
    \label{system_no}
\end{figure*}

Figure \ref{compact_types} shows the aggregate GW spectra of compact binaries from a stellar population with an initial mass of $10^{6} M_{\odot}$ formed in a single starburst at metallicity $Z = 0.020$, at different ages. For these plots and all subsequent ones, the distance between the observer and the stellar population is set as 1~kpc and the duration of observation is set as 4 years. Figure \ref{system_no} shows the number of systems in each frequency bin for each type of binary and age shown in Figure \ref{compact_types}. The frequency bins have a width of 0.2 dex. Figure \ref{compact_types} uses units of strain PSD, and the equivalent plots for characteristic strain are shown in Figure \ref{compact_types_hc}. Alternative arrangements of Figures \ref{compact_types} and \ref{system_no}, with each panel showing an individual age as opposed to an individual binary type, are shown in Figures \ref{compact_types_alt} and \ref{system_no_alt}, respectively.

General trends over age can be seen for each of the binary types: within the LISA frequency band from approximately $10^{-4}$ to $10^0$ Hz the strain increases until a peak and gradually decreases after that, while at lower frequencies the evolution of the spectrum is less regular but for most plots the strain is still increasing at $10^{11}$ years.

The age at which strain PSD in the LISA band (or characteristic strain overall) is maximal depends of the types of objects in the binaries, occurring later for the less massive remnants: it is at around $10^7$ years for the BH–BH and BH–NS binaries, at $10^8$ years for the BH–WD, NS–NS and NS–WD binaries and at $10^{10}$ years for WD–WD binaries. No WD–WD binaries have yet formed in the earliest time bin at $10^7$ years, while BH–WD and NS–WD binaries occur over a limited range of frequencies at this age.

This pattern with respect to binary types is as expected from stellar evolution, given that for NSs there is a greater delay time from stellar birth until the formation of the compact remnant than for BHs, and longer still for WDs.

For the majority of the binary types, the frequency at which the peak strain occurs shifts lower in frequency as the population ages. This is because for the binaries formed by at least one SN, new systems are not formed after about $\sim$100~Myr. Thus, the shortest-period binaries evolve out of the population, shifting the peak to lower frequencies. The exception is for the WD–WD binaries (which do not undergo SN), for which there is no clear trend. For these, the population is continually replenished by new WDs, so the evolution is less obvious.

The relative strain power of the different binary types is fairly even at most of the shown ages, with one major exception. The BH–BH and BH–NS binaries have maximal strain PSD values between $10^{-37}$ and $10^{-36}$ Hz$^{-1}$, while the BH–WD, NS–NS and WD–WD are somewhat higher at most ages, with maximal strain of around $10^{-35}$ Hz$^{-1}$ each. However, the NS–WD binaries (panel E in Figure \ref{compact_types}) have a significantly higher peak value that is between $10^{-34}$ and $10^{-33}$ Hz$^{-1}$, and from $10^8$ years onward they have the highest strain PSD in the LISA frequency range (and highest characteristic strain overall) until $10^{10}$ years, after which the WD–WD binaries have a comparable strain.

At $10^7$ years, the NS–WD binaries already have a high strain, but it is limited to a narrow frequency range around $10^{-4}$ Hz. This property is shared with the BH–WD binaries, which is something we will return to later in this section. After $10^8$ years, the rate of decrease of strain in the LISA frequency band is faster for the NS–WD and BH–WD than other types of binaries.

The discrepancy between the NS–WD binaries and the other types persists if, instead of looking at the GW strain, we look at the raw number of predicted systems in each frequency bin. This indicates that it is not caused by the NS–WD systems being treated differently to other binaries by the GW strain calculation code, but rather originates from the \textsc{bpass} population that is used as input. In Figure \ref{system_no}, we see that in the LISA frequency range only the WD–WD binaries (panel F) are more numerous, and only at the $10^{10}$ and $10^{11}$ year time bins, while from $10^8$ years onward the number of NS–WD binaries (panel E) exceeds that of any of the remaining four types by at least two orders of magnitude. The NS–WD binaries are also second highest in total number across the whole modelled frequency range, behind the WD–WD.

There are various pathways in the \textsc{bpass} population synthesis that produce these BH–WD and NS–WD binaries. The broad range of orbital frequencies in the population comes from binary systems where the primary star forms a NS in a tight orbit. The remnant then interacts with its companion while the latter is still on the main sequence. This leads to stable mass transfer, which causes little change in the orbit. However for some cases we find that enough mass transfer occurs to produce a BH by accretion-induced collapse of the NS. However these tend to only lead to orbits with lower frequencies up to $10^{-6}$~Hz but as we can see there is a significantly higher peak at $10^{-4}$~Hz.

This higher frequency peak comes from an unusual pathway. These arise from binary systems where the primary star forms a NS or BH at core collapse. In some of these systems significant mass is transferred to the companion and the systems are difficult to unbind with the first supernova kick at the companions have effective initial masses from 20 to 120~M$_{\odot}$. The orbit after the first core collapse also has a relatively tight initial orbital period from 1 to 4 days. The stellar wind mass-loss rates at the more massive end of this distribution are significant and the mass of the star changes at a similar rate as the nuclear evolution in the core. This leads to the stellar core being exposed relatively rapidly. At this point, and for the initially less massive stars, Roche lobe overflow and in many cases common-envelope evolution also occurs, further increasing the mass loss rates. This leads the stellar model to rapidly evolve towards a white dwarf in the very tight orbits at the peak we seek.

During this common-envelope evolution our BPASS prescription leads to the initially massive star evolving to a WD, orbiting a BH or NS that has not accreted significant mass in a very short orbit. These systems would effectively appear as ultra-luminous or low-mass X-ray binaries with short period orbits, depending on which phase of evolution they are in. At the end of evolution the binary has a BH that is still relatively close to its initial mass (between 3 to 7~M$_{\odot}$) in an extremely short period of around $10^{-4}$~Hz. In some cases the companion quickly becomes a low-mass star that has already experienced significant hydrogen fusion but may continue to experience nuclear burning on a timescale longer than the orbital evolution by gravitational waves.

This peak is metallicity-dependent because at lower metallicities the mass range at which this evolution occurs shifts to lower initial masses, removing the very massive stars, and so the mass range of stars that leads to this evolution moves into the 10 to 25~M$_{\odot}$ range. As a result, those binaries appear at later times and more spread out in time compared to the sharp peak at higher metallicities.

This evolutionary path to form NS/BH–WD binaries is somewhat uncertain and may just be a spurious evolutionary pathway due to the implementation of stellar winds, mass transfer and common-envelope evolution within BPASS. It does however reveal that new and important LISA sources may occur that are not predicted by other binary population synthesis codes. These stars will also be difficult to observe electromagnetically as they are faint, sub-solar stripped stars orbiting around a BH in orbital periods of hours. However, there are known NS–WD binaries within the Galaxy (e.g. \citet{knownnswd1,knownnswd2}; catalogued in e.g. \citet{atnf_pulsar_catalogue}). We also note that \citet{ulx2} predict that 40\% of NS–WDs in the LISA band have as progenitors X-ray binaries with super-Eddington accretion that could be detectable in EM as ultra-luminous X-ray sources \citep{ulx1}.

Of course the large population of these predicted binaries might be only possible in the \textsc{bpass} binary evolution algorithm and may not be robust, but even if LISA reveals this pathway does not exist it is important to be prepared for sources in the eventual LISA data that we have not yet predicted.

We note that the intermittent empty bins in the NS–WD and WD–WD spectra (panels E and F of Figures \ref{compact_types} and \ref{system_no}) are not physical in origin, but an artefact resulting from the \textsc{bpass} data set not containing any systems to populate those bins because of the finite resolution of the grid of initial system parameters.

\subsection{Spectra of living vs. compact binaries}

\begin{figure*}
    \centering
    \includegraphics[width=1\columnwidth]{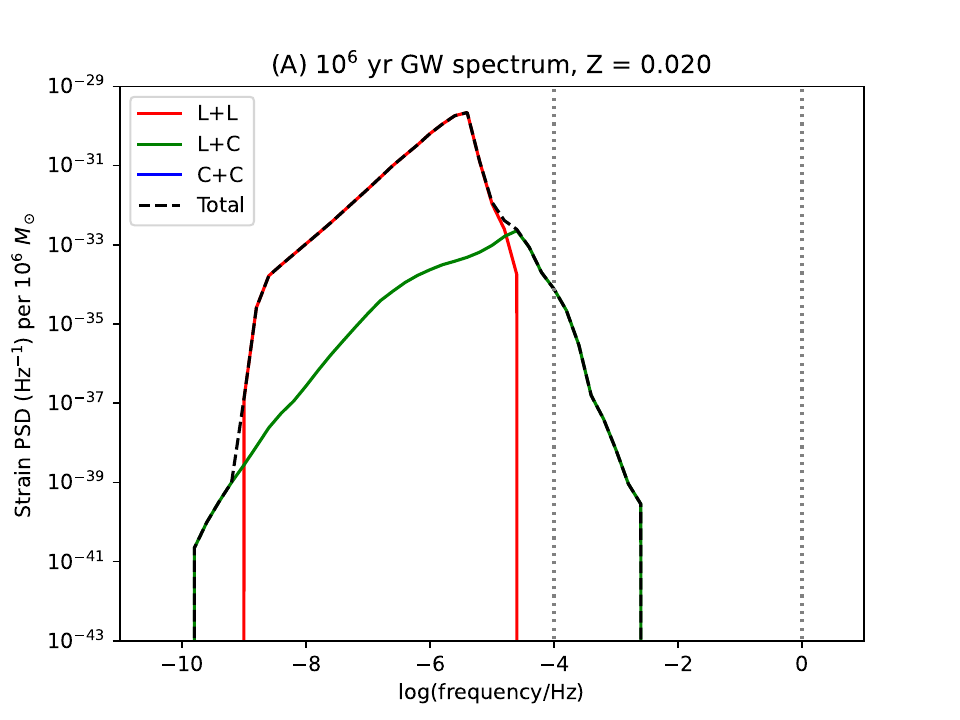}
    \includegraphics[width=1\columnwidth]{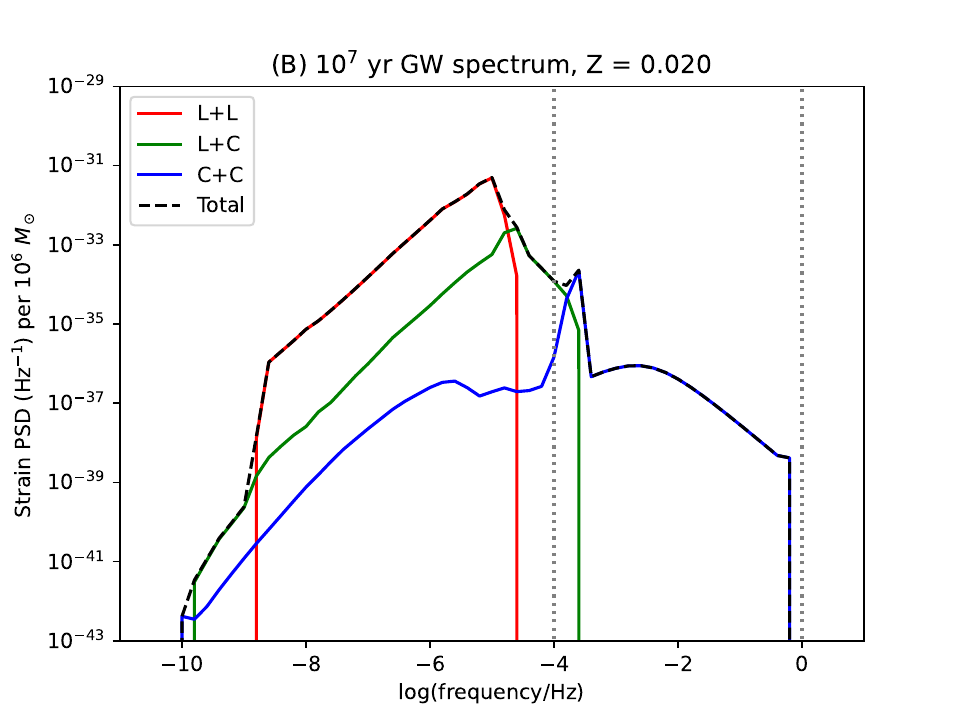}
    \includegraphics[width=1\columnwidth]{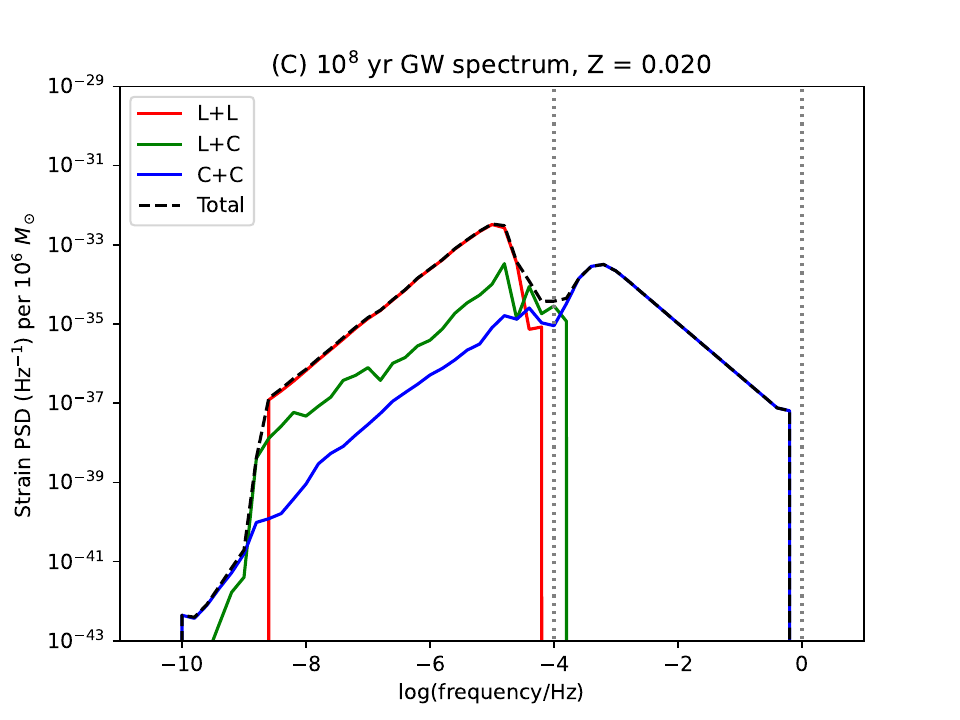}
    \includegraphics[width=1\columnwidth]{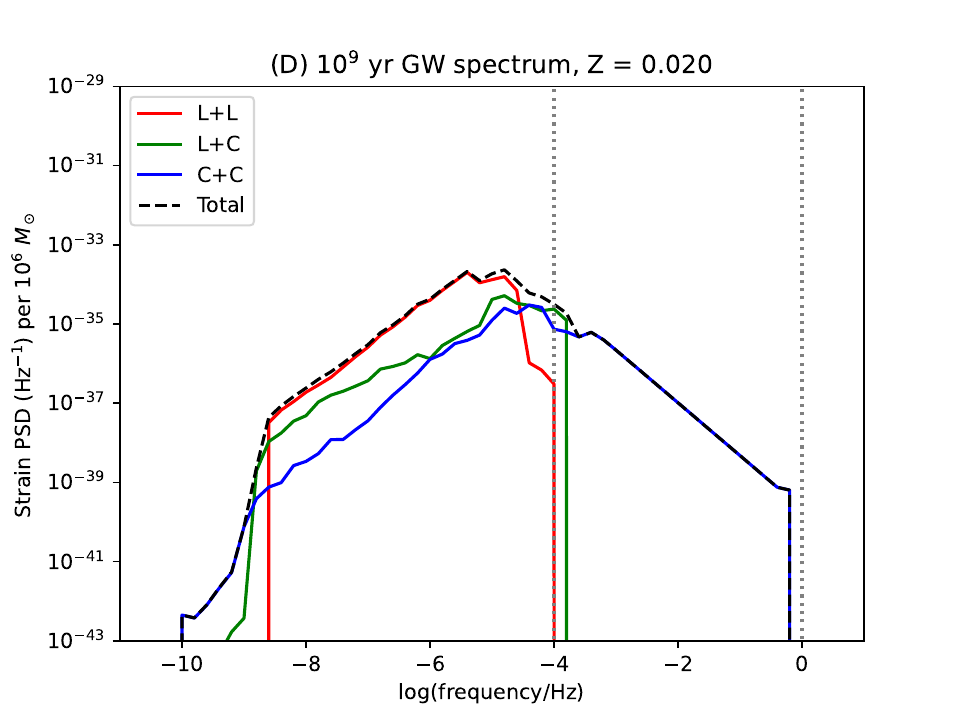}
    \includegraphics[width=1\columnwidth]{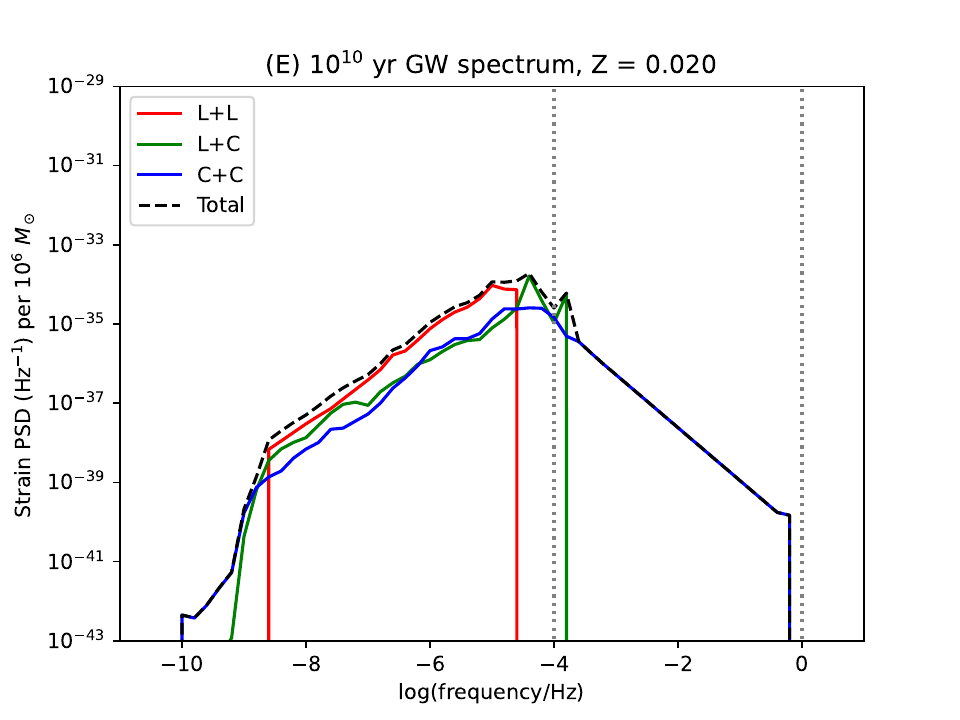}
    \caption{Gravitational wave spectra of living and compact binaries for a stellar population with an initial mass of $10^6$ M$_{\odot}$ at a distance of 1 kpc and $Z$ = 0.020, at different ages. ``L+L'' are binaries of two living stars, ``L+C'' are binaries with one living star and one compact remnant and ``C+C'' are double compact binaries. The black dashed line shows the total spectrum of the population at each age. The dotted lines show an approximation of the LISA frequency range.}
    \label{living_compact}
\end{figure*}

Figure \ref{living_compact} compares the spectra of all compact binary types added together, those of binaries of two living stars and those of one living star and one compact remnant, for the same stellar population as Figure \ref{compact_types}. Also plotted is the total gravitational wave spectrum of the population, consisting of the living–living, living–compact and compact–compact spectra added together. Each subplot shows the spectra at a different age, from $10^6$ to $10^{10}$ years. Figure \ref{living_compact} uses units of strain PSD, and the equivalent plots for characteristic strain are shown in Figure \ref{living_compact_hc}.

We note that the living–compact binaries with periods less than a day would be interacting binaries that may be observable electromagnetically. BH–WD and NS–WD binaries at sub-minute periods may also be interacting due to the WD filling its Roche lobe, but we have not included this in our modelling.

The spectra of the living and compact binaries are clearly distinct: the strain power of the living binaries is largely within the range of $10^{-10}$ to $10^{-5}$ Hz, the living–compact binaries are similar but extend to somewhat higher frequencies at $10^{-4}$ Hz, while compact binaries are found over the entire modelled frequency range up to $10^0$ Hz as discussed in section \ref{sec_compact}. In the characteristic strain plots, one can see distinct peaks formed by the living binaries between $10^{-6}$ and $10^{-5}$ Hz and the compact binaries between $10^{-4}$ and $10^{-2}$ Hz, though the latter is less prominent in the strain PSD plots.

The reason that living binaries are not found beyond the cutoff around $10^{-5}$ to $10^{-4}$ Hz is that, at higher frequencies, the orbital radius would be sufficiently small that the stars' envelopes would interact with each other, the end result of which would be either the two stars merging or the orbit being widened due to mass loss.

The spectra of the living and compact binaries are also different in terms of their evolution over time: the living binaries have their highest strain at $10^6$ years, an age at which no double compact binaries have even formed yet, and consistently decrease in strain over time, while the compact binaries form later and have comparable strain power (and greater characteristic strain) from around $10^8$ to $10^9$ years. This pattern is as expected based on the evolution of living stars to compact remnants.

We note that the living–compact binaries' maximum frequency of $10^{-4}$~Hz matches the peak of the NS–WD and BH–WD binaries we discussed above. This indicates the peak in those populations is a result of these binaries, which would likely to be observed as low-mass X-ray binaries before the companion becomes a WD.

The highest strain power in Figure \ref{living_compact} overall is for the living binaries at the youngest ages (Panel A), with a strain power between $10^{-30}$ and $10^{-29}$ Hz$^{-1}$ (characteristic strain between $10^{-18}$ and $10^{-17}$) at $10^6$ years. However, this peak occurs at $10^{-6}$ Hz, a frequency too low for LISA to detect. The highest strain of double compact binaries is between $10^{-34}$ and $10^{-33}$ Hz$^{-1}$ (characteristic strain between $10^{-19}$ and $10^{-18}$) from $10^8$ to $10^9$ years (panels C and D), which corresponds to the predominant NS–WD binaries discussed in section \ref{sec_compact}.

Because the compact and living binaries peak at different frequencies and which of these classes is predominant changes as the population grows older, the overall GW spectrum of the stellar population changes over time. We investigate this in more detail in section \ref{sec_evol}.

\subsection{Evolution of spectra over time and for different metallicities} \label{sec_evol}

\begin{figure*}
    \centering
    \includegraphics[width=1\columnwidth]{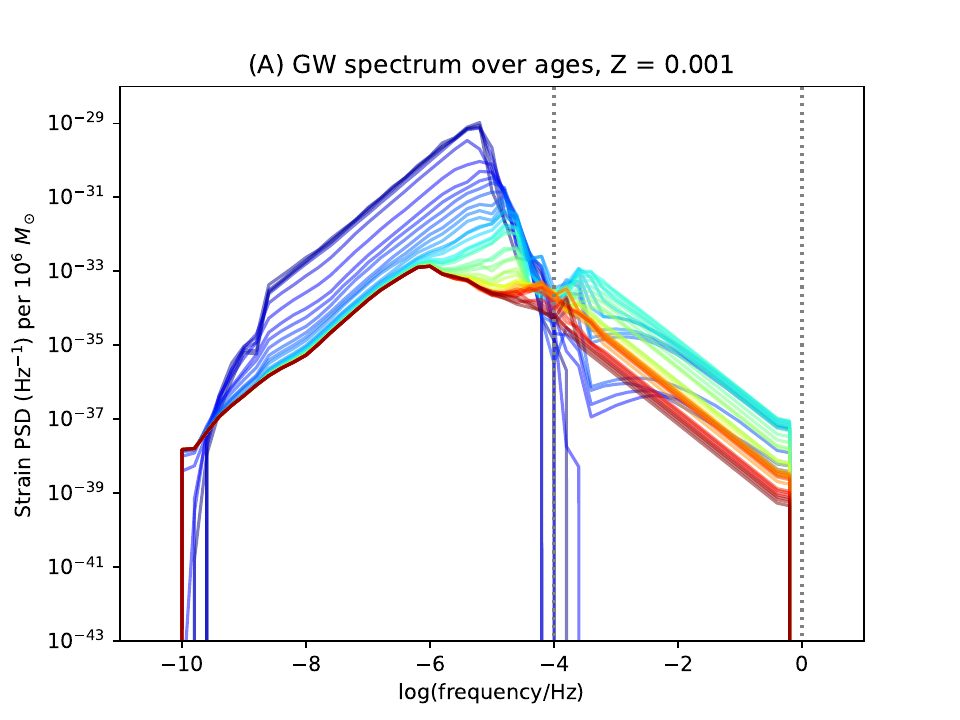}
    \includegraphics[width=1\columnwidth]{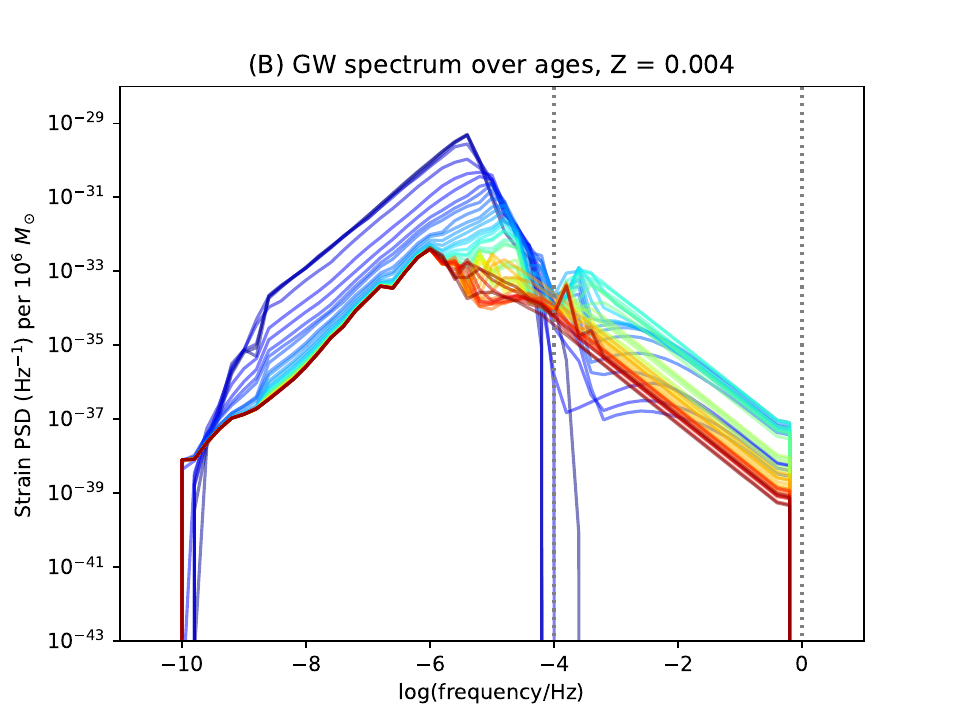}
    \includegraphics[width=1\columnwidth]{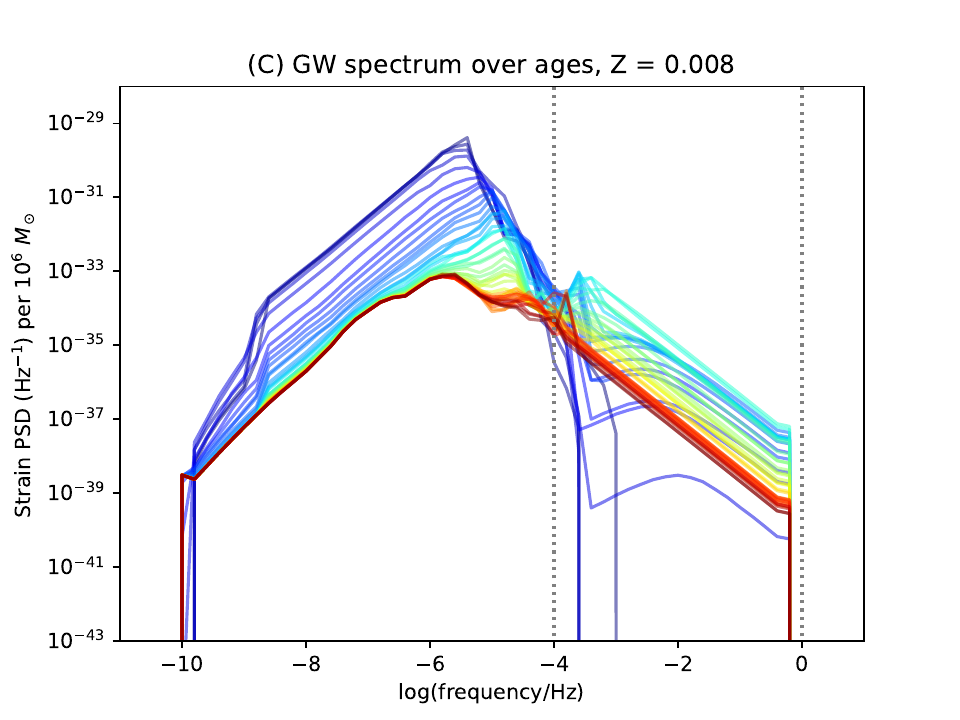}
    \includegraphics[width=1\columnwidth]{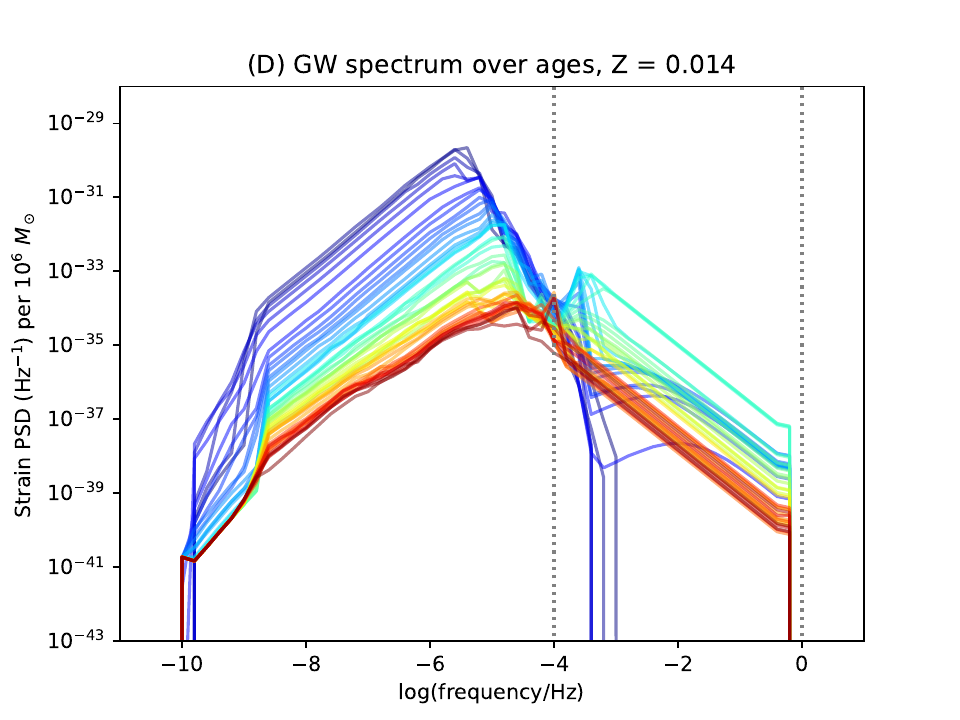}
    \raisebox{-0.5\height}{\includegraphics[width=1\columnwidth]{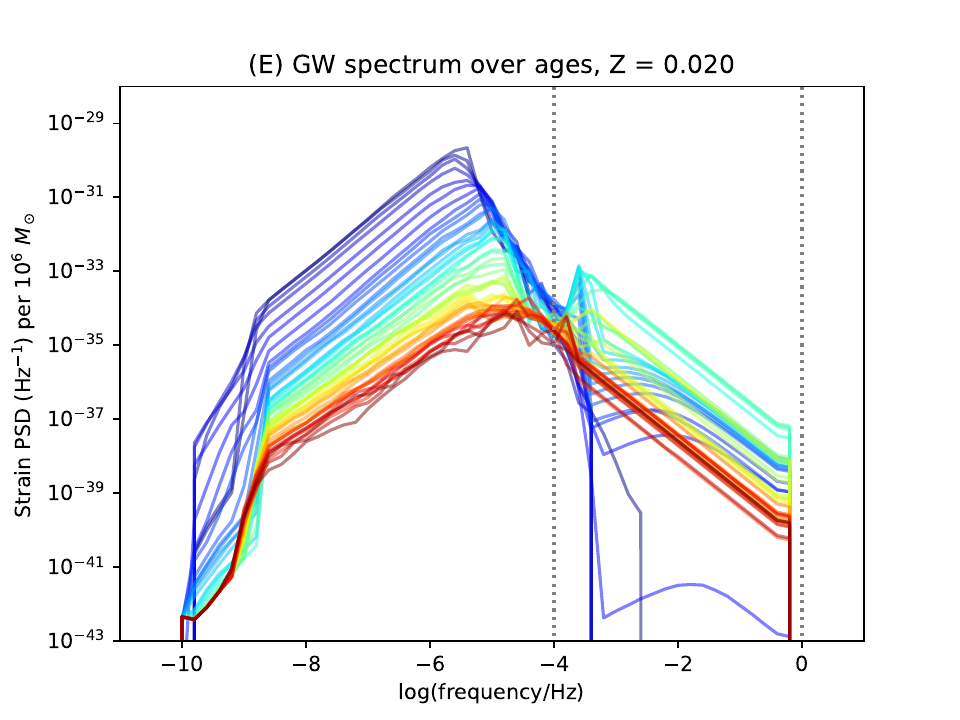}}
    \raisebox{-0.5\height}{\includegraphics[height=4.5cm]{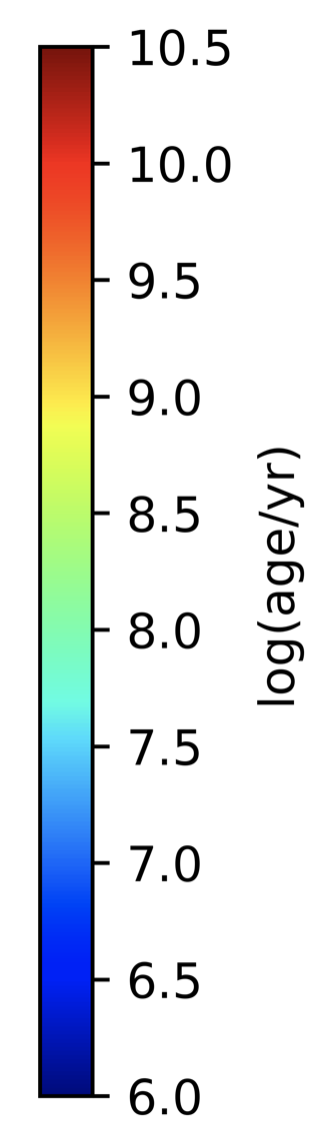}}
    \caption{Total GW spectrum of a stellar population with an initial mass of $10^6$ M$_{\odot}$ at a distance of 1 kpc, at ages from $10^6$ yr (blue) to $10^{10.5}$ yr (red) in steps of 0.1 dex, for several different metallicities. The dotted lines show an approximation of the LISA frequency range.}
    \label{evol_age}
\end{figure*}

Figure \ref{evol_age} shows the total (living and compact) GW spectrum of a $10^6 M_{\odot}$ stellar population over a range of ages: each line is a different age, going from $10^6$ years (blue) to $10^{10.5}$ years (red) in steps of 0.1 dex. Each subplot shows the spectra at a different metallicity, from $Z = 0.001$ to $Z = 0.020$. Figure \ref{evol_age} uses units of strain PSD, and the equivalent plots for characteristic strain are shown in Figure \ref{evol_age_hc}.

The high resolution in age allows us to see the changes in the spectrum over time more clearly. For all metallicities we can see a peak between $10^{-6}$ and $10^{-5}$ Hz that has its highest amplitude when the population is youngest and loses prominence over time, while there is a broad region of strain in the LISA frequency range between $10^{-4}$ Hz to $10^0$ Hz which appears between $10^7$ and $10^8$ years, rapidly rising to a maximum strain and then slowly declining.

The shape of the spectrum is similar for all metallicities, but there are several differences that can be observed. Firstly, the strain below the LISA frequency range (dominated by living binaries) is higher for lower metallicities (panels A and B), though there is little effect on the strain power within the LISA frequency range.

Another distinction is that, at higher metallicities (panels D and E), we can see the appearance of a sharp peak at $10^{-4}$ Hz at around $10^7$ years, which exists for several time bins before spreading out into the broad strain across the LISA frequency range. This relates to the formation of NS–WD and BH–WD binaries discussed in section \ref{sec_compact}. The sharp peak consists of these binaries as they are initially formed, and these are then smeared out over higher frequencies as they evolve under gravitational radiation.

At low metallicities, we do not observe a sharp peak at $10^{-4}$ Hz prior to the appearance of the broad region of strain from compact binaries, as described in section \ref{sec_compact}. However, a bump does appear in the spectrum at the same frequency much later, after $10^{10}$ years, which is not observed at higher metallicities.

\subsection{Comparing spectra to LISA sensitivity}

\begin{figure*}
    \centering
    \includegraphics[width=1\columnwidth]{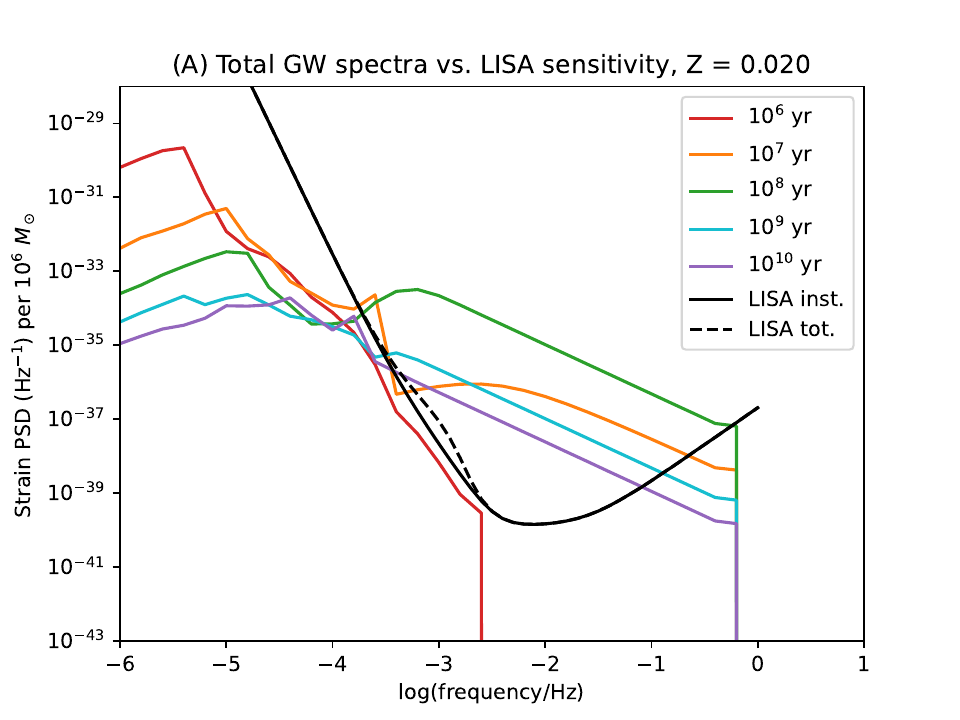}
    \includegraphics[width=1\columnwidth]{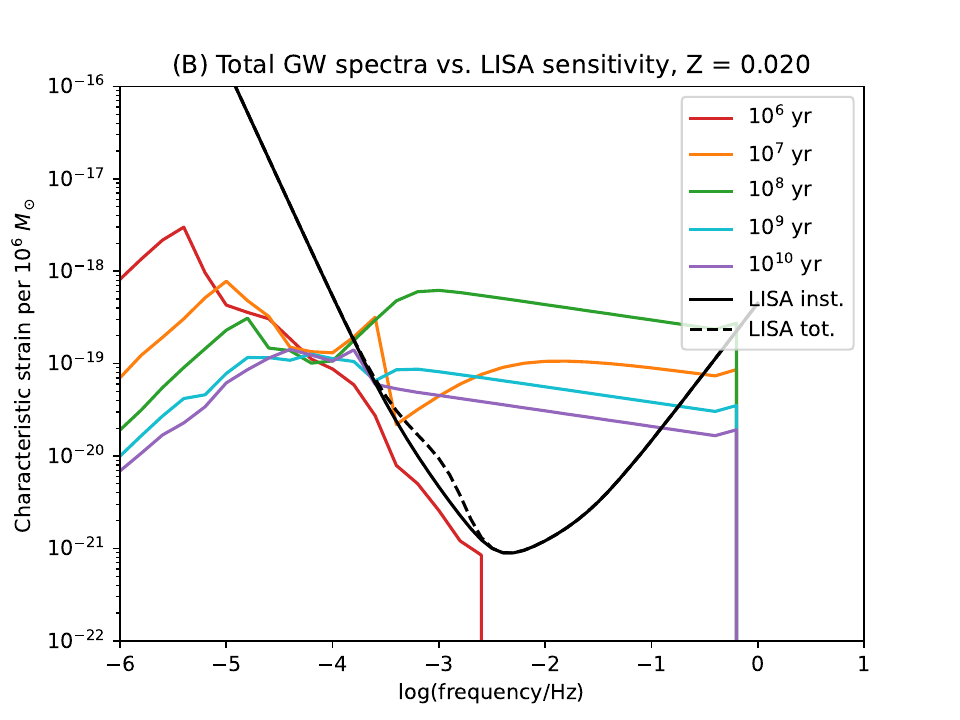}
    \caption{Comparison of total GW spectrum at different ages to LISA sensitivity curve, for a stellar population with an initial mass of $10^6$ M$_{\odot}$ at a distance of 1 kpc and $Z$ = 0.020, in PSD and characteristic strain units.}
    \label{lisa_comparison}
\end{figure*}

Figure \ref{lisa_comparison} shows the total (living and compact) GW spectrum of a $10^6 M_{\odot}$ stellar population at 1 kpc, at several different ages, compared to the LISA sensitivity curve. The sensitivity curve is shown in two versions, one with only instrumental noise (equation \ref{sc_instrument}) and one with added confusion noise for four years of observation (equation \ref{sc_confusion}). The two subplots show the strain PSD and characteristic strain, respectively.

It can be seen that, from $10^7$ years onward, the spectrum lies above the sensitivity curve from a frequency of roughly $10^{-3.8}$ Hz, up to between $10^{-1}$ and $10^0$ Hz depending on the age. At $10^{-2}$ Hz, around the maximum sensitivity of LISA, the spectra are several orders of magnitude above the noise, for both methods of quantifying the strain. This suggests that this stellar population would be clearly detectable by LISA from $10^7$ years onward.

However, that statement has a major caveat: as can be seen in Figure \ref{system_no}, for frequencies above around $10^{-4}$ Hz, the expected number of systems in each frequency bin is relatively low, with the number per bin being less than one for each of the binary types, and at the highest frequencies several orders of magnitude less.

\begin{table*}
    \centering
    \begin{tabular}{ c | c c c c c c c }
    Age (yr) & BH–BH & BH–NS & BH–WD & NS–NS & NS–WD & WD–WD & Total \\
    \hline
       \multicolumn{8}{c}{Z = 0.020 for LISA sensitivity}  \\
    \hline
    $10^7$ & $2.5 \times 10^{-4}$ & $8.0 \times 10^{-4}$ & $0.015$ & $6.5 \times 10^{-4}$ & $4.7$ & $0$ & $4.7$ \\
    $10^8$ & $3.0 \times 10^{-4}$ & $3.0 \times 10^{-3}$ & $0.29$ & $0.11$ & $14$ & $0$ & $15$ \\
    $10^9$ & $1.4 \times 10^{-4}$ & $1.4 \times 10^{-3}$ & $6.3 \times 10^{-3}$ & $0.046$ & $1.6$ & $0$ & $1.6$ \\
    $10^{10}$ & $4.5 \times 10^{-5}$ & $3.2 \times 10^{-4}$ & $4.4 \times 10^{-4}$ & $8.9 \times 10^{-3}$ & $0.033$ & $2.6$ & $2.7$ \\
    $10^{11}$ & $4.5 \times 10^{-6}$ & $2.4 \times 10^{-5}$ & $0$ & $4.7 \times 10^{-4}$ & $0.021$ & $0.73$ & $0.75$\\
   \hline
       \multicolumn{8}{c}{Z = 0.001 for LISA sensitivity}  \\
    \hline
    $10^7$ & $2.5 \times 10^{-4}$ & $2.2 \times 10^{-4}$ & $1.9$ & $2.6 \times 10^{-11}$ & $0$ & $0$ & $1.9$ \\
    $10^8$ & $4.5 \times 10^{-4}$ & $0.083$ & $4.9$ & $0.23$ & $34$ & $0$ & $39$ \\
    $10^9$ & $2.8 \times 10^{-4}$ & $0.032$ & $0.43$ & $0.087$ & $2.2$ & $14$ & $17$ \\
    $10^{10}$ & $9.2 \times 10^{-5}$ & $6.2 \times 10^{-3}$ & $0.11$ & $0.015$ & $0.052$ & $2.8$ & $3.0$ \\
    $10^{11}$ & $2.2 \times 10^{-5}$ & $3.8 \times 10^{-4}$ & $1.7 \times 10^{-3}$ & $6.7 \times 10^{-4}$ & $0.012$ & $2.3$ & $2.3$\\
   \hline
       \multicolumn{8}{c}{Z = 0.020 for $\mu$Ares sensitivity} \\
    \hline
    $10^7$ & $1.2 \times 10^{-3}$ & $0.013$ & $0.71$ & $0.010$ & $5.1$ & $0$ & $5.9$ \\
    $10^8$ & $2.7 \times 10^{-3}$ & $0.094$ & $0.66$ & $3.1$ & $26$ & $20$ & $50$ \\
    $10^9$ & $3.8 \times 10^{-3}$ & $0.079$ & $0.27$ & $2.3$ & $14$ & $1.7 \times 10^2$ & $1.8 \times 10^2$ \\
    $10^{10}$ & $4.3 \times 10^{-3}$ & $0.044$ & $0.079$ & $1.1$ & $18$ & $2.5 \times 10^2$ & $2.7 \times 10^2$ \\
    $10^{11}$ & $1.1 \times 10^{-3}$ & $6.8 \times 10^{-3}$ & $0.011$ & $0.11$ & $1.4$ & $2.4 \times 10^3$ & $2.4 \times 10^3$\\
   \hline
       \multicolumn{8}{c}{Z = 0.001 for $\mu$Ares sensitivity}  \\
    \hline
   $10^7$ & $1.2 \times 10^{-3}$ & $3.9 \times 10^{-3}$ & $4.3$ & $6.7 \times 10^{-11}$ & $0$ & $0$ & $4.3$ \\
    $10^8$ & $9.0 \times 10^{-3}$ & $2.8$ & $17$ & $6.1$ & $49$ & $21$ & $96$ \\
    $10^9$ & $0.017$ & $2.3$ & $12$ & $4.5$ & $67$ & $3.0 \times 10^2$ & $3.9 \times 10^2$ \\
    $10^{10}$ & $0.022$ & $1.2$ & $7.4$ & $1.9$ & $19$ & $2.8 \times 10^2$ & $3.1 \times 10^2$ \\
    $10^{11}$ & $9.3 \times 10^{-3}$ & $0.13$ & $0.87$ & $0.16$ & $1.1$ & $6.0 \times 10^2$ & $6.0 \times 10^2$\\

    \end{tabular}
    \caption{Number of binaries of each remnant type within two specific frequency ranges, at several ages, for a stellar population of $10^6$ M$_\odot$ at 1 kpc and two different metallicities.}
    \label{number_table}
\end{table*}

\begin{figure*}
    \centering
    \includegraphics[width=1\columnwidth]{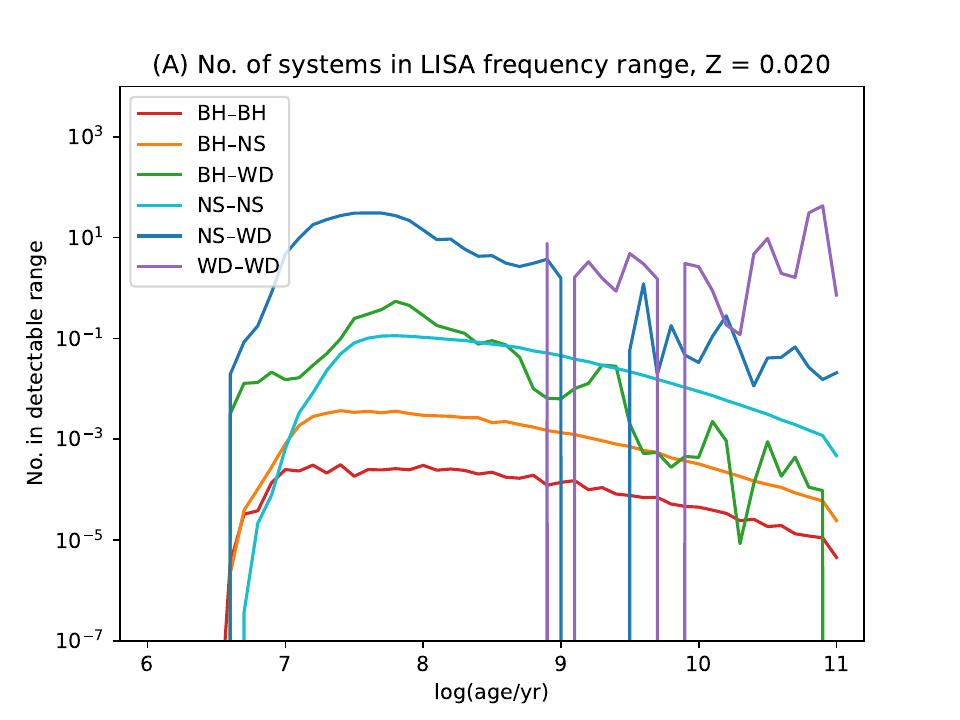}
    \includegraphics[width=1\columnwidth]{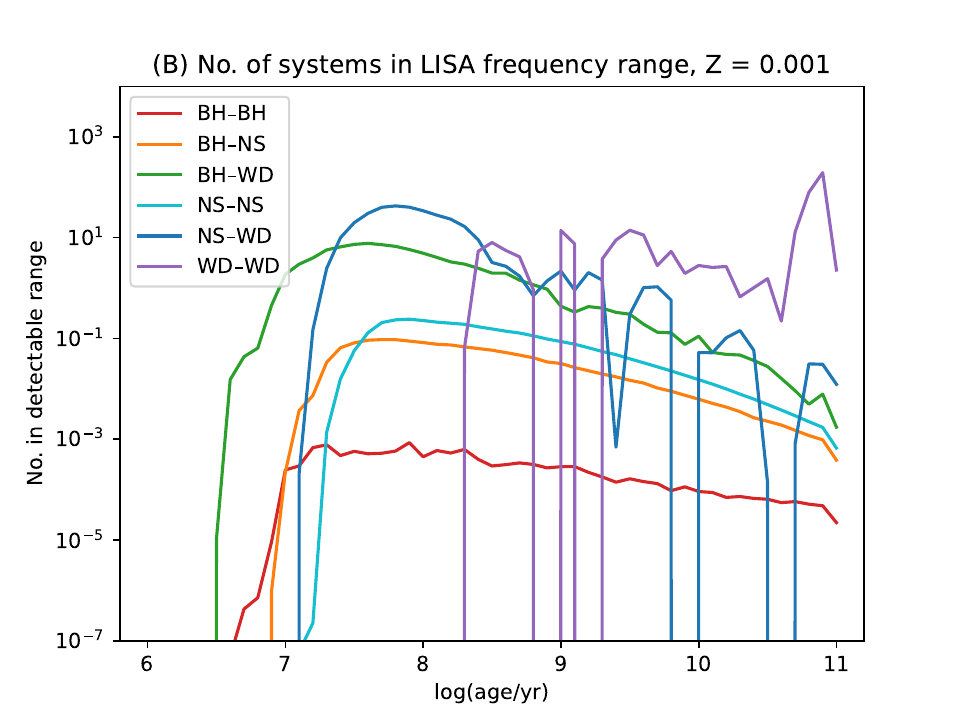}
    \includegraphics[width=1\columnwidth]{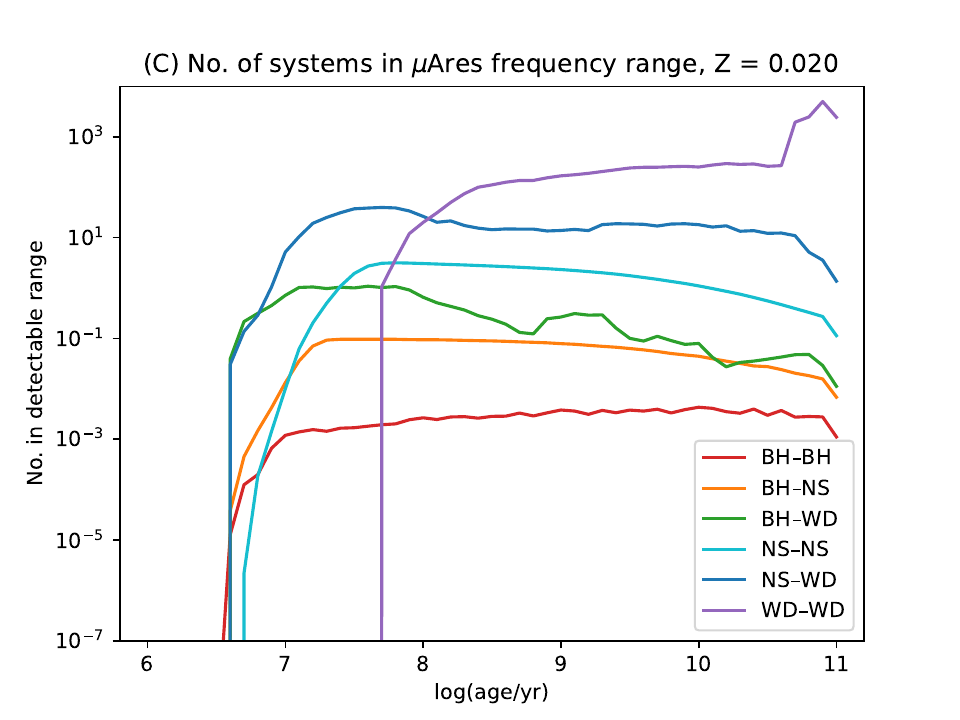}
    \includegraphics[width=1\columnwidth]{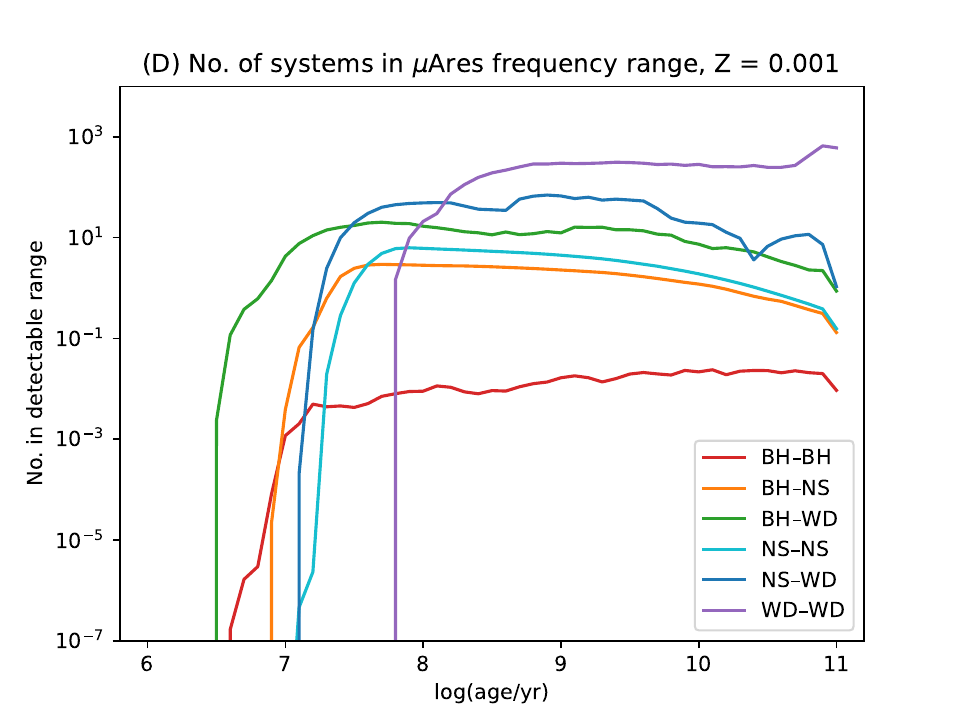}
    \caption{Number of binaries of each remnant type within two specific frequency ranges, at several ages, for a stellar population of $10^6$ M$_\odot$ at 1 kpc and two different metallicities, as in Table \ref{number_table}.}
    \label{number_fig}
\end{figure*}

To illustrate this in more detail, Table \ref{number_table} shows a rough estimate of the number of binaries of each type that would be detectable by LISA over a range of ages and two different metallicities, $Z = 0.020$ and $Z = 0.001$. This estimate was made by summing the number of systems in each frequency bin down to $10^{-3.8}$ Hz, this frequency being used as an approximation of the lowest frequency where the GW spectra exceed the LISA noise. The same data is also shown graphically (with more intermediate ages) in Figure \ref{number_fig}.

Using this approximation of LISA sensitivity, we see that, for $Z = 0.020$ (panel A), the total number of detectable systems is greater than 1 for all ages except $10^{11}$ years, but for the individual types only NS–WD and WD–WD binaries have a predicted number greater than 1 in any time bin. For $Z = 0.001$ (panel B), the total numbers exceed those of $Z = 0.020$ for all ages except $10^7$ years, and the total is greater than 1 for all ages. At this metallicity, the number of BH–WD binaries also exceeds 1 in some time bins. It is worth noting that at $Z = 0.001$, the BH–WD binaries are mostly within one order of magnitude of the NS–WDs in number, unlike at $Z = 0.020$ where they are two orders of magnitude less frequent than the NS–WDs at most ages.

Something that can be learned from these relatively low total numbers is that, for a globular cluster which may have an initial mass of around $10^6$ M$_\odot$, the individual spectrum of the cluster in the LISA frequency range likely would not be smooth as in Figure \ref{lisa_comparison}, but rather consist of a number of distinct peaks, which would be above the averaged curve and so probably individually detectable. However, for an object like the Large Magellanic Cloud (LMC), with a stellar mass on the order of $10^9$ to $10^{10}$ M$_\odot$ \citep{lmcstellarmass1,lmcstellarmass2} and a distance of approx. 50 kpc \citep{lmcdistance}, the spectrum may approach a smooth curve, although the LMC is sufficiently large on the sky that it might be possible to resolve individual sources within it \citep{roebber2020,korol2020,lmclisa}. Note that strain of a population is proportional to square root of the number of binaries (and therefore the square root of the total mass) and inversely proportional to the distance, so assuming the mass and distance stated previously the strain of the LMC would be comparable to that of our population.

\subsection{Comparing spectra to \textit{µ}Ares sensitivity}

For comparison, in Table \ref{number_table} and Figure \ref{number_fig} we also calculate the numbers of detectable systems for the proposed detector $\mu$Ares \citep{muares}. The frequency range that $\mu$Ares could observe overlaps with that of LISA but extends down to around $10^{-6}$ Hz; for our calculation we have made a conservative estimate and taken $\mu$Ares as being able to detect GWs to one order of magnitude in frequency lower than LISA, i.e. to $10^{-4.8}$ Hz.

The total number of detectable systems in our approximation of the frequency range of $\mu$Ares is close to that of LISA in the earliest time bins, but grows more quickly over time (as expected because the range is wider) and by $10^9$ years is more than an order of magnitude greater. For $Z = 0.020$ (panel C), for NS–NS, NS–WD and WD–WD binaries have a predicted number greater than 1 for $\mu$Ares, and for $Z = 0.001$ (panel D) this is true for all remnant binary types except for BH–BHs. Note also that $\mu$Ares might just be able to detect the peak of living binaries around $10^{-6}$ Hz, but we have included only double compact binaries in this analysis.

As in Figures \ref{compact_types} and \ref{system_no}, the intermittent empty bins in Figure \ref{number_fig} are a result of the finite resolution of the grid of initial system parameters in \textsc{bpass}, which causes there to be some time bins where there are no models of a certain type that are formed at that time or carried over from the preceding time bin.

\section{Galactic clusters as likely sources for LISA} \label{chapter_examples}

In this section we calculate the expected number of binaries in the LISA frequency range for catalogued open and globular clusters, to illustrate the implications of the results in section \ref{chapter_results} for real clusters.

\subsection{Open clusters}

Open clusters are highly suitable as real-world examples to illustrate our calculations, as due to their relatively low density compared to globular clusters, dynamical interactions do not have a significant effect upon their stellar populations. To calculate the expected number of binaries in the LISA frequency range – defined, as in Table \ref{number_table}, as those binaries with frequencies greater than $10^{-3.8}$ Hz (and less than $10^0$ Hz) – we need to consider their age, metallicity and initial mass. Note that, as in Table \ref{number_table}, this is only a calculation of how many binaries are present within the frequency range, and does not take into account the distance of the clusters or evaluate the detectability of the binaries.

We perform these calculations for the objects in two catalogues of Galactic open clusters, \citet{catalogue_piskunov} and \citet{catalogue_cordoni}. These catalogues provide the current mass of each cluster while we require the initial mass, so we compute the latter from the former using \textsc{bpass} surviving stellar mass files \citep{bpass1,bpass2}. These files give the total stellar mass at different ages given an initial mass of $10^6$ M$_{\odot}$ for the stellar population. Additionally, \citet{catalogue_piskunov} does not specify the metallicities of its clusters, so we assume that these have a metallicity close to solar (Z = 0.014) as they are in the Galactic disk.

In Table \ref{open_cluster_table}, we present the data for each of the clusters in \citet{catalogue_piskunov} and \citet{catalogue_cordoni} that has an expected number of binaries of at least 0.1.

From these results, we can see that open clusters do not tend to have high numbers of binaries in the LISA frequency due to their relatively low mass. Only one of the open clusters in the two catalogues has an expected number of binaries greater than one. However, the sum of the expected numbers for the nine clusters in Table \ref{open_cluster_table} is 3.09, which indicates that, even though most of these clusters are individually unlikely to contain binaries, when considering them as an ensemble it is probable that a few of these clusters will have binaries. Across the 236 clusters in \citet{catalogue_piskunov} for which the mass is given, there are 4.23 expected binaries in total, and for the 78 clusters in \citet{catalogue_cordoni} the total is 1.18 (note that there are some clusters which appear in both catalogues, though none of these have more than 0.1 expected binaries).

The majority of the open clusters that have at least 0.1 binaries are between 10 and 100 Myr old. Younger clusters will not have formed many compact binaries yet (as shown in e.g. Figure \ref{number_fig}) and older open clusters tend to lose stars to cluster evaporation. Given these ages, we can see by comparison to the data in e.g. Figure \ref{number_fig} or Table \ref{number_table} that any binaries that do appear in open clusters would be most likely NS–WD binaries, or possibly BH–WD.

One particularly interesting cluster amongst those listed in Table \ref{open_cluster_table} is Melotte 20 (also known as the Alpha Persei Cluster), which has the third-highest expected number of binaries at 0.31 and is significantly closer to Earth than most of the clusters at 190 pc, so if this clusters does contain a binary in the LISA frequency range, it is quite likely to be detectable.

We note that our list of open clusters based on \citet{catalogue_piskunov} and \citet{catalogue_cordoni} is non-exhaustive, and as more clusters are found by Gaia our understanding of where GW sources may exist will increase. For example, there are known young clusters with large numbers of red supergiants that suggest they could be massive, but as they are behind large amounts of extinction towards the Galactic center their total mass and age are uncertain \citep[e.g][]{bpassrsgclusters}.

\subsection{Globular clusters}

For globular clusters, as opposed to open clusters, there are effects of dynamical interactions that would affect the stellar population as it evolves. Therefore, our \textsc{bpass} model populations based upon isolated binary evolution would be less accurate for globular clusters than for open clusters. However, for globular clusters \textsc{bpass} can still be used as a first-order approximation as long as we remain aware of these uncertainties.

In Table \ref{globular_cluster_table}, we present the date for the 20 most massive clusters located within 10 kpc of Earth from the catalogue of \citet{catalogue_baumgardt1} and \citet{catalogue_baumgardt2} and references therein.

Compared to the open clusters, we see that the globular clusters on average contain more binaries in the LISA frequency range, due to their higher masses and older ages. 14 clusters have at least one expected binary, and six have at least two. $\omega$ Centauri, the most massive of the clusters in our sample, has approx. 12 expected binaries. An observation of multiple binary GW sources from the same cluster would aid in identifying these sources as being associated with the cluster.

The total number of expected binaries in the LISA frequency across the 20 clusters in Table \ref{globular_cluster_table} is 42.5. As these globular clusters are all at least $10^{10}$ years old, we can see by comparison to the data in e.g. Figure \ref{number_fig} or Table \ref{number_table} that these would be mostly WD–WD binaries.

\section{Discussion and conclusions} \label{chapter_discussion}

We have presented here a first attempt at producing a synthetic gravitational wave spectrum for a simple stellar population. As in any research we have had to make caveats and simplifications in our work. First, our results depend on the predictions of the \textsc{bpass} binary stellar evolution models. These have been verified against many observations of stellar systems \citep{wofford2016,bpass1,bpass2,bpassccsn,bpasstransient,bpassblap,bpassrsg}. However there are clear indications that the models are not complete and need revision \citep{bpassmasstransfer}. There is also physics missing for some of the closest binaries: for example, magnetic wind braking \citep{magnetic_wind_braking,mwb1,mwb2,mwb3} is not yet included in the evolution of WD binaries. Therefore our WD–WD source population may need to be revised to have more short period binaries.

We have also assumed throughout this work that the stellar binaries evolve isolated from each other and no dynamical interactions modify the binary star population. In reality, in the most massive and densest globular clusters dynamical interactions lead to a significant hardening of the binary population \citep{heggie1975,ivanova2006,ivanova2008,portegies2010}, as well as the destruction of some binaries. Our predicted number of sources could therefore vary on the order of a few as the result of such effects, which should be investigated in the future. However, other than this our results are relatively robust. With population synthesis there is always the risk that one spurious stellar model with a significant weight could create an erroneous prediction. We have checked our models and find all significant features are the result of 10s to 1000s of stellar models with moderate weights within the population. Thus, uncertainty in our results is dominated by the input physics of the stellar evolution and binary interactions, as well as ignoring the possible dynamical interactions.

We can see from Table \ref{number_table} and Figure \ref{number_fig} that different types of compact objects are predominant at different ages, such as NS–WD binaries between 10$^7$ and 10$^8$ years and WD–WD binaries after 10$^9$ years. One overall conclusion that we can draw is that the total number of expected GW sources in a star cluster peaks at the ages of 10$^8$ to 10$^9$ years. This implies that if a star cluster is identified to be a source of GWs, then this fact could provide an extra constraint on its age. Alternatively, if the age is known then the mass or metallicity of the cluster could be estimated.

Our predicted numbers of WD–WD binaries detectable by LISA is of the order of a few at late ages. It is difficult to assess whether this is an underestimate or not. There are several physical processes that are not currently include in the \textsc{bpass} population that may increase our predicted population if they were included. As previously mentioned, \textsc{bpass} does not yet simulate magnetic wind braking. Secondly, older stellar clusters are typically globular clusters, wherein dynamical interactions would increase the number of WD–WD binaries in the LISA frequency range. Finally, the greater stability of mass transfer in \textsc{bpass} and our unique treatment of common envelope events (CEE) \citep[for details, see][]{bpassmasstransfer} are uncertain and may increase or decrease the number of any compact remnant binaries by causing fewer or more stellar mergers, respectively.

There are no confirmed EM observations of BH–WD binaries, although these would be difficult to observe due to the faintness of the components. Accretion in BH–WDs could lead to X-ray emission, in particular that of ultracompact X-ray binaries (UCXBs) \citep{bhwd_ucxb}; the fact that BH–WDs have not been observed while UCXBs with NSs have \citep{nelemans_ucxb} could be because of the difficulty of classifying the components from EM observations alone, which would not be the case for GW observations \citep{sberna2021}. Particularly violent accretion that could occur in BH–WDs has also been proposed as a possible source of gamma-ray bursts \citep{bhwd_grb}.

However, as mentioned in section \ref{sec_compact}, there have been numerous confirmed EM observations of NS–WD binaries. For example both \citet{desvignes2016} and \citet{fonseca2016} detail a number of binaries that have a millisecond pulsar within a binary system. Nine of these have sub-day orbital periods, the shortest being for J0348+0432 which has a period of 2.45 hours; that is, a GW frequency of 0.23 mHz, which is at the peak frequency in our predictions for such systems. For systems where the period is greater than a day, the distribution extends up to periods of order of 150~days, thus covering the full range of frequencies we predict in Figure \ref{system_no}. This gives us confidence that, while uncertainties in our predictions remain, the general picture does match with expectations from observed NS–WD binaries. 

The mergers of BH–WD and NS–WD binaries may produce EM transients or sources, depending on the structure of the WD and the mass ratio of the two remnants. For example, the tidal disruption of a WD by a companion BH or NS could produce transients broadly similar to those of subluminous Type I supernovae \citep{metzger2012}. Or for the specific case of a BH–WD binary with an additional stellar-mass BH companion causing tidal disruption of the WD, high-energy X-ray transients or gamma-ray transients similar to long GRBs \citep{fragione2020} could be produced. These systems could also produce longer-lasting LMXBs \citep[e.g.][]{bahramian2017} or even become Thorne-Żytkow objects which may still have some detectable GW emission in the LIGO frequency range \citep{tzo_gw}.

We have also taken a sample of relatively nearby and well-studied massive star clusters in the Milky Way, predicting the number of GW sources within these from their masses, ages and metallicities where available. For younger (open) clusters we find that only a small fraction have a high chance of containing GW transients, with those with the greatest number generally being between 10$^7$ to 10$^8$ years in age. The chance of a source being in each individual cluster is low but over all there is a high probability that at least three of the open clusters we examined should have GW sources.

Comparatively, for the older globular clusters we investigated, each in our list is more likely to contain one to several sources because of their significant masses. We estimate these predictions are within an order of magnitude of the number of sources expected, because of the dynamical processes that occur within globular clusters. However, our results still indicate that these systems may harbour large numbers of GW sources detectable by LISA.

We have plotted the sky location of these clusters in Figure \ref{sky_location}. As we would expected, the open clusters lie primarily in the galactic plane, while the globular clusters are scattered more widely, though centred on the Galactic centre. LISA's sensitivity response function is complex, but as it orbits about the Sun it will on average be most sensitive approximately in the plane of the ecliptic. This is plotted in Figure \ref{sky_location} and we see that a grouping of clusters lie within this expected most sensitive region and are thus prime targets for analysis of the LISA signal.

This suggests that studying clusters with EM and GW observations together will allow strong constraints on binary evolution. Here we predict the GW signals that could be seen; for clusters of known mass and age we could compare their observed GW spectra from LISA to these predictions. If the observed GWs were to be different from the predictions, then either our understanding of binary evolution is wrong, or dynamical effects in clusters dominate isolated binary evolution. Future work will be to create a list of clusters in the Milky Way and Magellanic Clouds where this method could be employed. We note another reason to use known clusters is that, as shown by \citet{finch2023}, knowing the sky locations of potential GW sources can significantly aid data analysis of the LISA signal.

This work is part of the broader project within the \textsc{bpass} collaboration to understand gravitational wave sources \citep[e.g.][]{bpassmassdist,bpassmasstransfer,bpasskick}. This article is the first where we have begun to predict the sources expected to be observed by LISA. While here we have concentrated on individual star clusters in Tang et al. (in prep) we have utilized these same \textsc{bpass} predictions with a model of the Milky Way Galaxy to make a prediction for the total source population. Given that \textsc{bpass} has several features that set it apart from other binary population synthesis codes, most importantly following the evolution of the stars in a detailed stellar evolution code, the predictions we present and those in future provide an important contribution to the the growing work on understanding the GW sky for future GW observatories.

In summary, we draw the following conclusions:
\begin{enumerate}
    \item Various types of GW sources contribute to the spectrum of a simple stellar population like that of a stellar cluster. The relative strengths of these sources can vary depending on the population's age and metallicity.
    \item The most populous sources are WD–WD binaries, followed by BH/NS–WDs. Both of these categories could contribute O(10-100) sources for a initial stellar population of $10^6$ M$_{\odot}$, albeit at different times in the evolution. Other remnant types are less populous (less than 1 expected for $10^6$ M$_{\odot}$) and similar in number; of these, NS–NS are the most populous, although these will depend on the magnitude of NS natal kicks.
    \item The sources in the LISA band are mostly NS–WD binaries at young ages, and later WD–WDs.
    \item BH/NS–WD systems may be important sources of GWs. Within \textsc{bpass} a formation pathway including CEE during early evolution of the BH/NS binary leads to formation of a low mass WD around the BH/NS in a short orbit that will be a significant source of GWs.
    \item There are also many living–compact remnant binaries, which could be X-ray bright, that may be observed by LISA. The study of such systems is important as they may be strong LISA GW sources even before both stars become compact remnants.
    \item Metallicity has a significant impact on binary evolution and thus the eventual GW sources, with different types of binaries being more or less dominant depending on metallicity, and overall more sources being formed at low metallicities.
    \item Individual clusters could have their age and mass constrained by GW observations, especially for clusters with ages around $10^8$ to $10^9$ years, and masses over $10^5$ M$_{\odot}$.
    \item Using the information known about star clusters within our Galaxy will allow the identification of sky locations where one could expect to find GW sources that LISA may observe.
    \item While LISA will concentrate on the double compact remnant binaries, $\mu$Ares will potentially find more of the living–compact remnant systems and living binaries.
\end{enumerate}

\section*{Acknowledgements}

WGJvZ acknowledges support from Radboud University, the European Research Council (ERC) under the European Union’s Horizon 2020 research and innovation programme (grant agreement No.~725246), Dutch Research Council grant 639.043.514 and the University of Auckland. JJE acknowledges support from Marsden Fund Council grant MFP-UOA2131 managed through the Royal Society of New Zealand Te Apārangi. PNT acknowledges support from the University of Auckland. The authors thank Renate Meyer, Gijs Nelemans, Simon F. Portegies Zwart and Elizabeth R. Stanway for useful discussions.

\section*{Data Availability}

The \textsc{bpass} data set can be downloaded from \url{https://bpass.auckland.ac.nz}. Other data underlying this article will be shared on reasonable request to the corresponding author.



\bibliographystyle{mnras}
\bibliography{references} 




\appendix

\section{Characteristic strain plots}

Included in this appendix are alternative versions of several plots in the main body of the paper, with GW strain quantified in units of characteristic strain rather than strain PSD.

\begin{figure*}
    \centering
    \includegraphics[width=1\columnwidth]{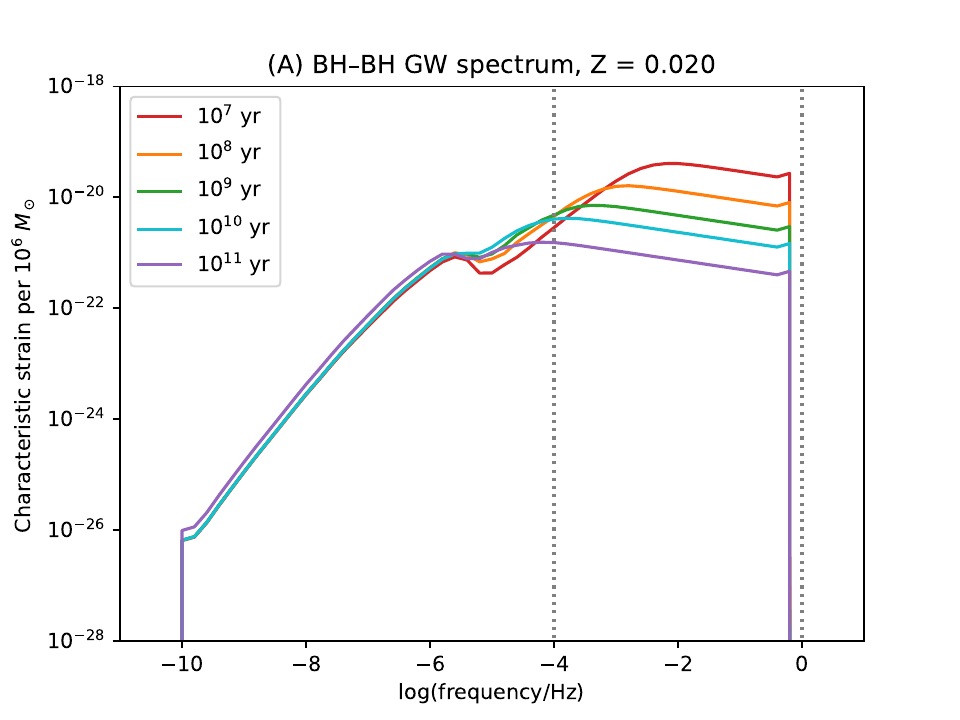}
    \includegraphics[width=1\columnwidth]{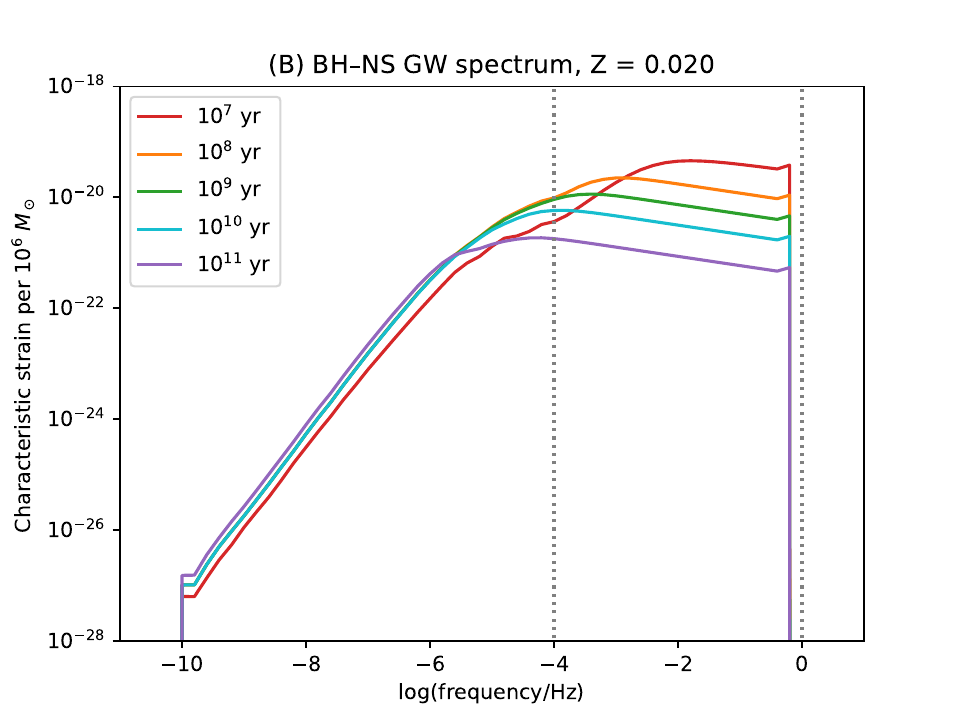}
    \includegraphics[width=1\columnwidth]{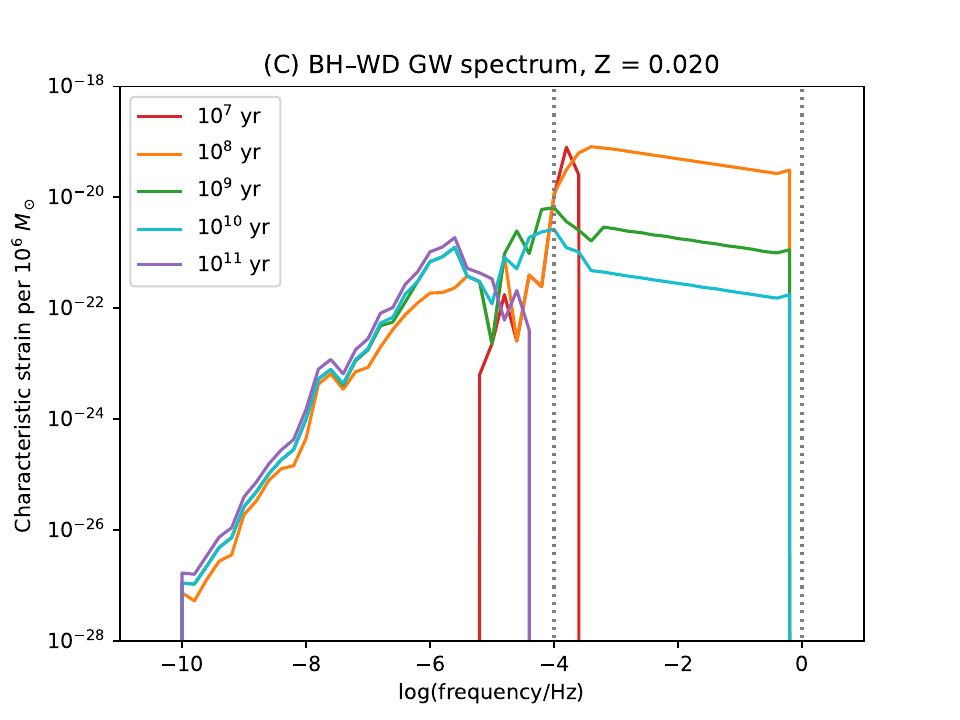}
    \includegraphics[width=1\columnwidth]{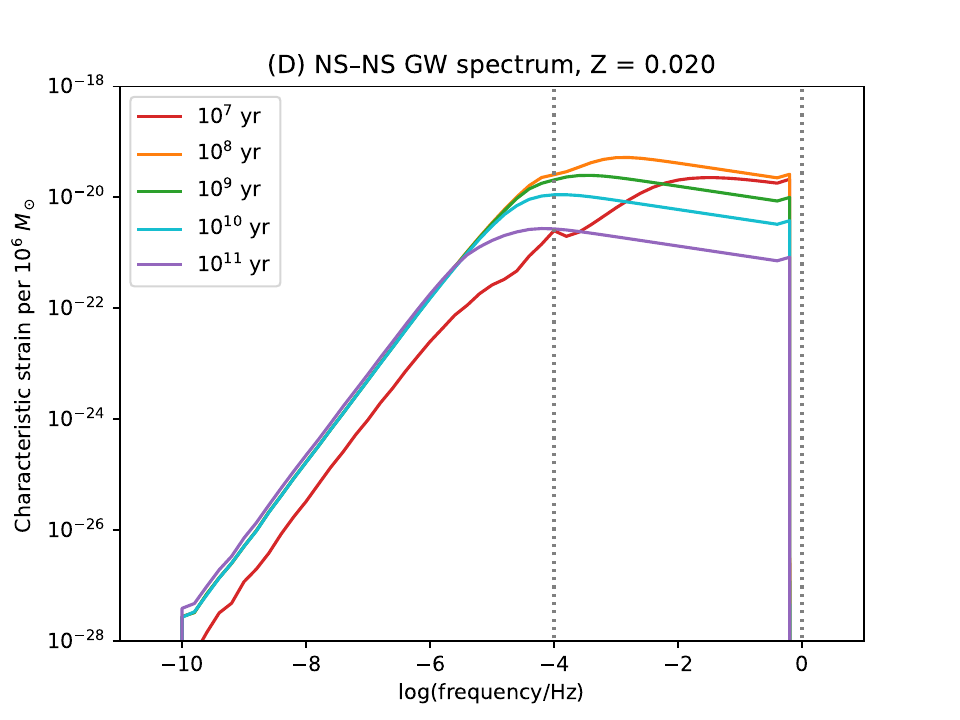}
    \includegraphics[width=1\columnwidth]{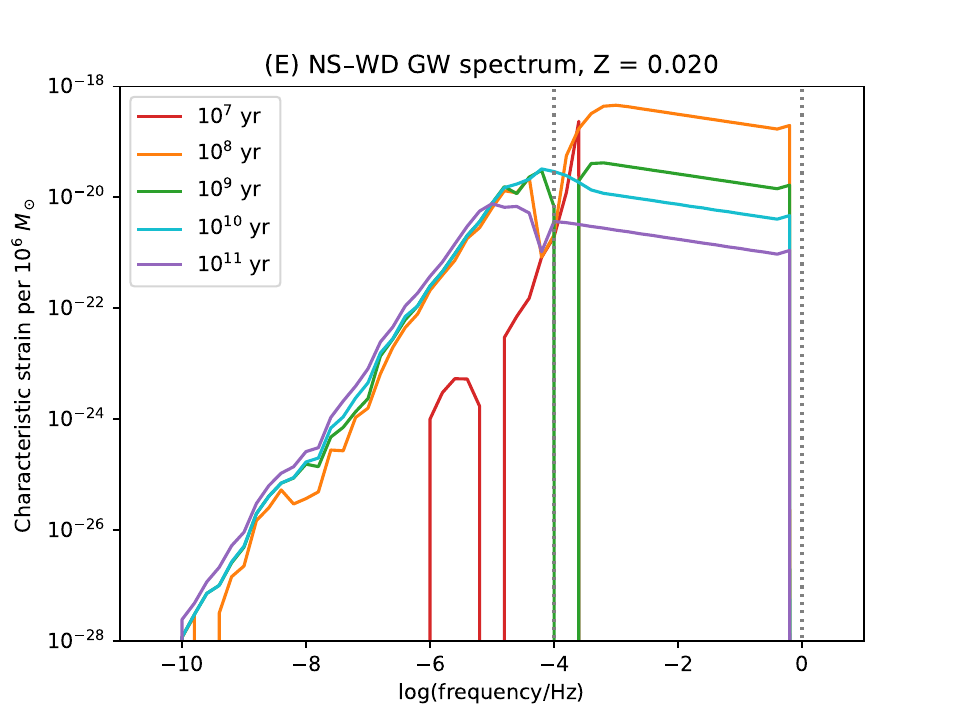}
    \includegraphics[width=1\columnwidth]{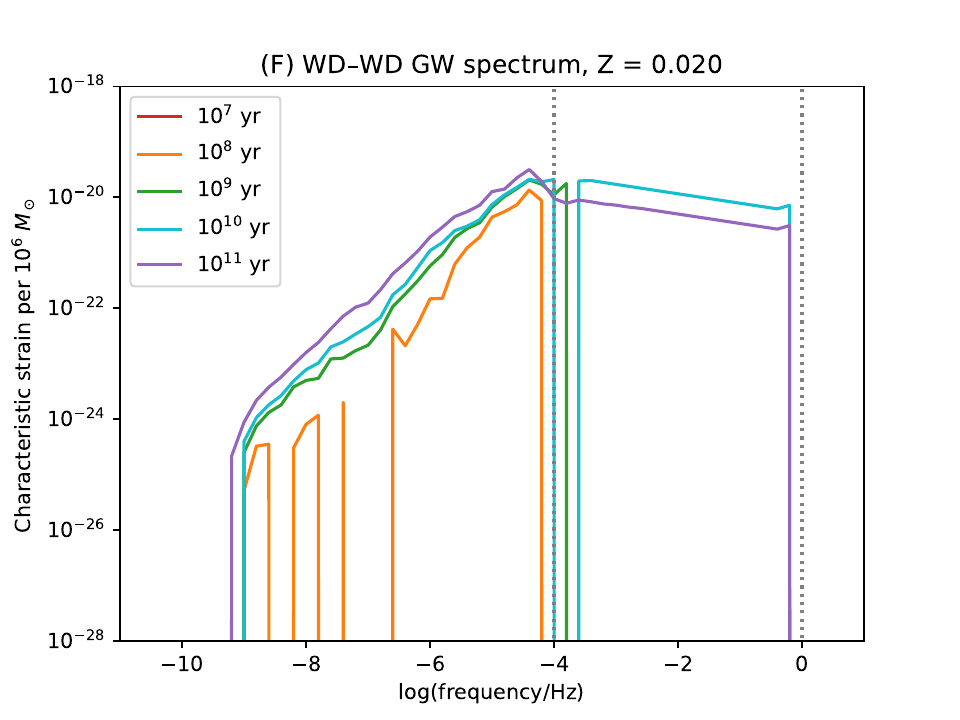}
    \caption{Same as Figure \ref{compact_types}, but with characteristic strain plotted instead of strain PSD.}
    \label{compact_types_hc}
\end{figure*}

\begin{figure*}
    \centering
    \includegraphics[width=1\columnwidth]{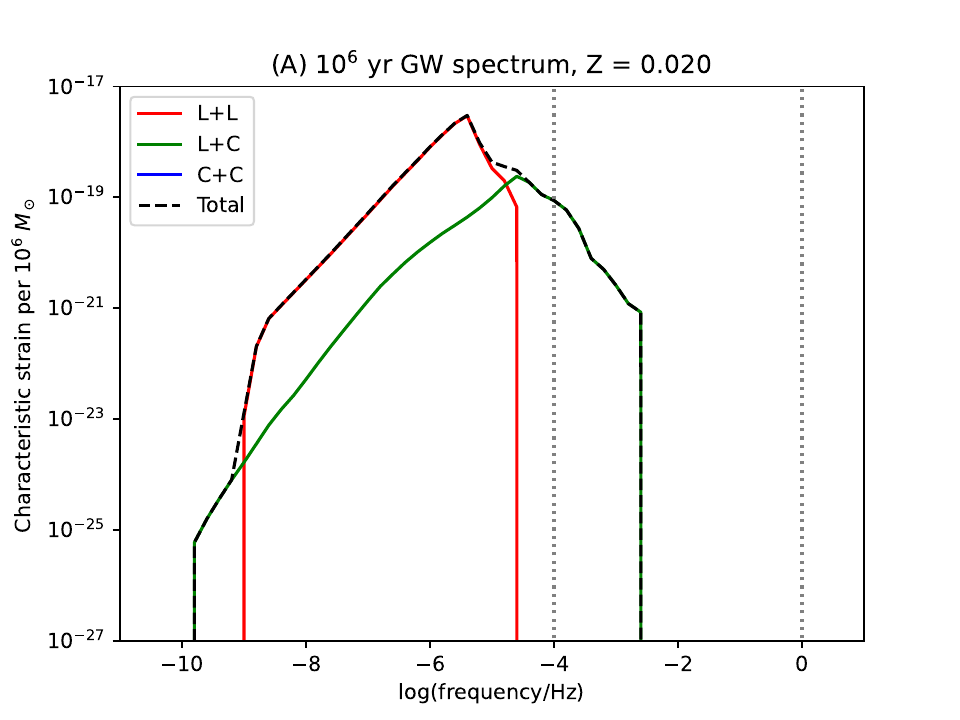}
    \includegraphics[width=1\columnwidth]{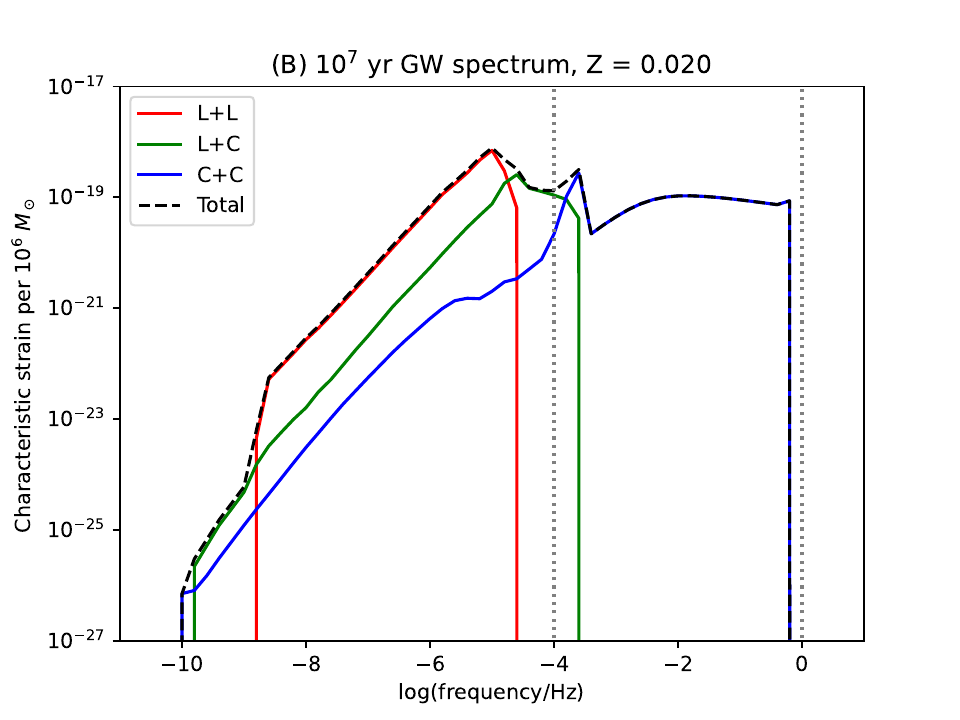}
    \includegraphics[width=1\columnwidth]{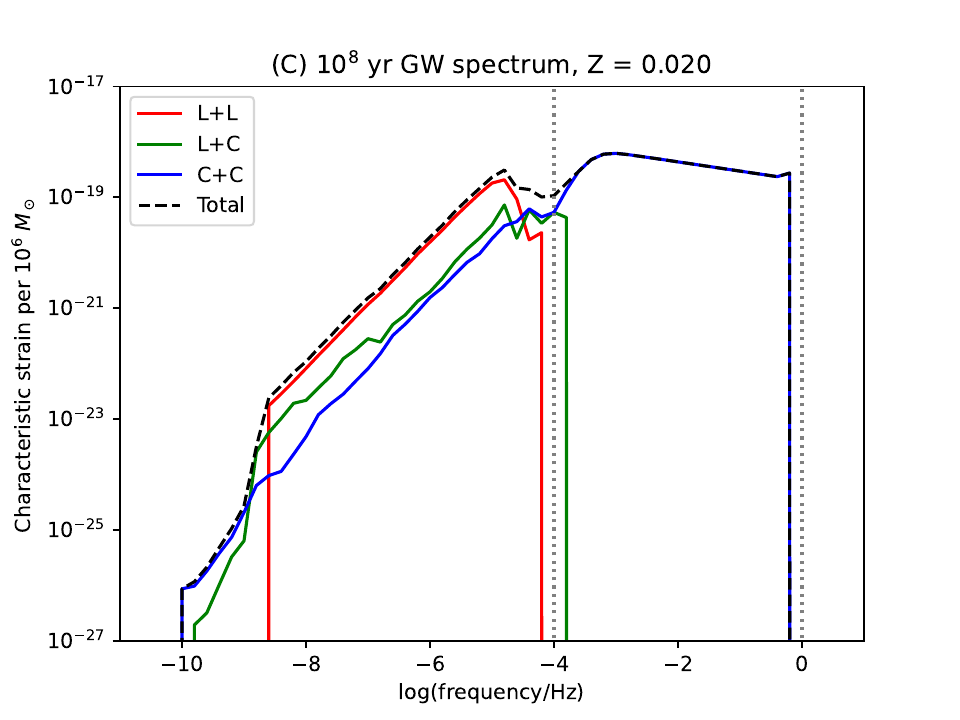}
    \includegraphics[width=1\columnwidth]{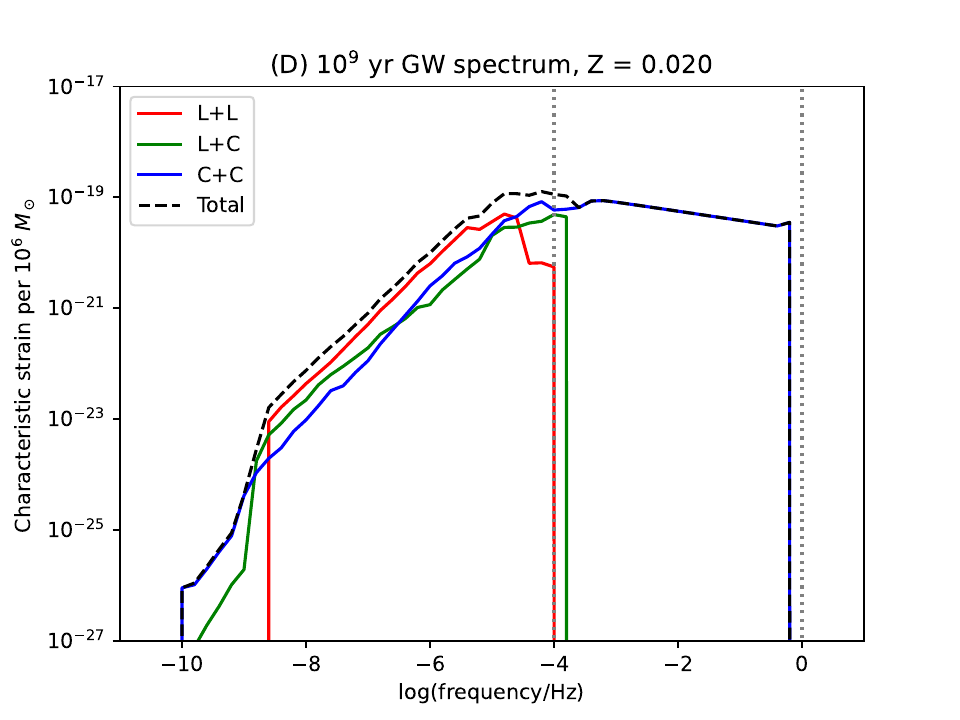}
    \includegraphics[width=1\columnwidth]{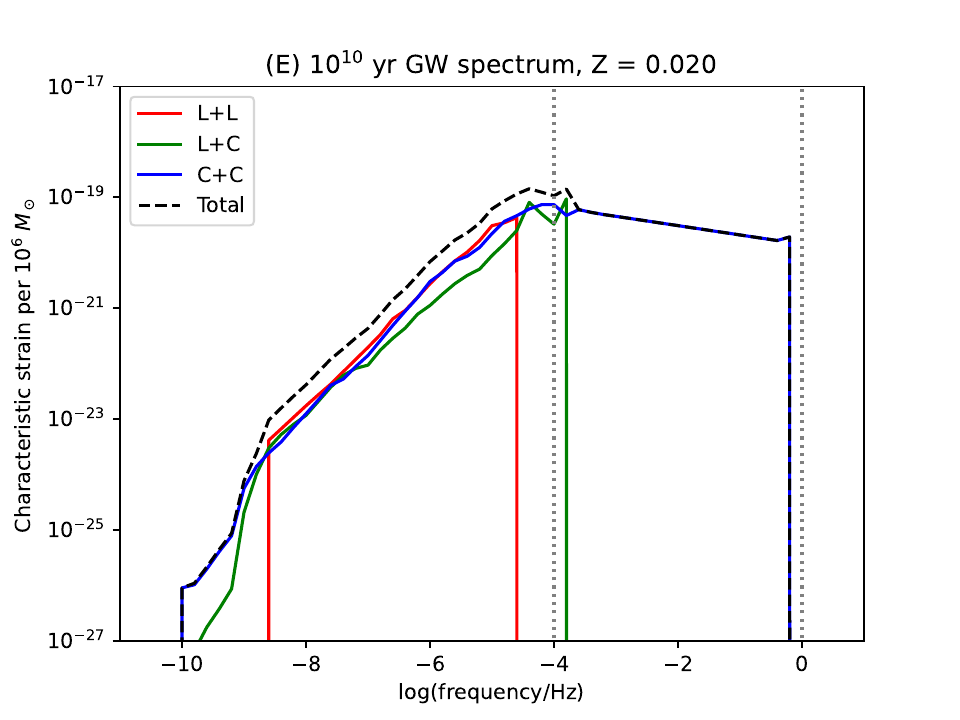}
    \caption{Same as Figure \ref{living_compact}, but with characteristic strain plotted instead of strain PSD.}
    \label{living_compact_hc}
\end{figure*}

\begin{figure*}
    \centering
    \includegraphics[width=1\columnwidth]{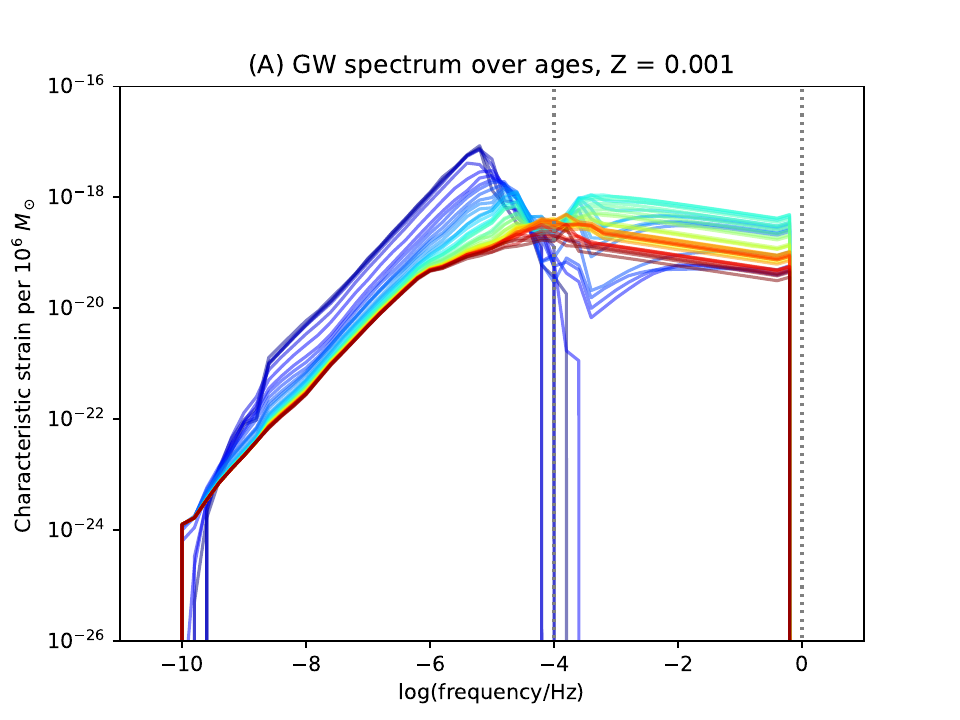}
    \includegraphics[width=1\columnwidth]{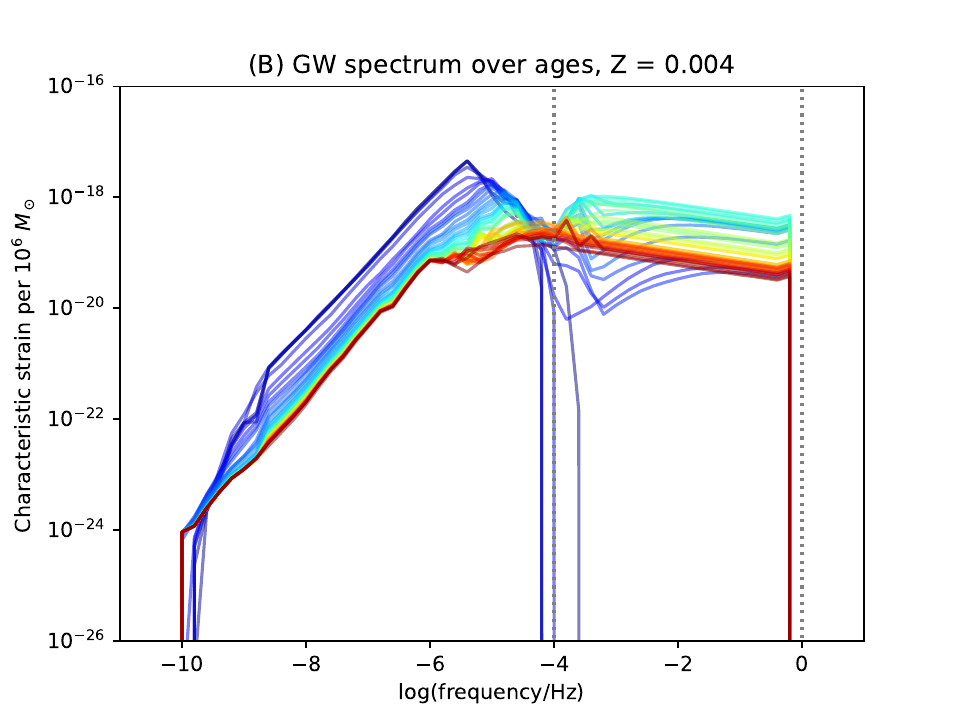}
    \includegraphics[width=1\columnwidth]{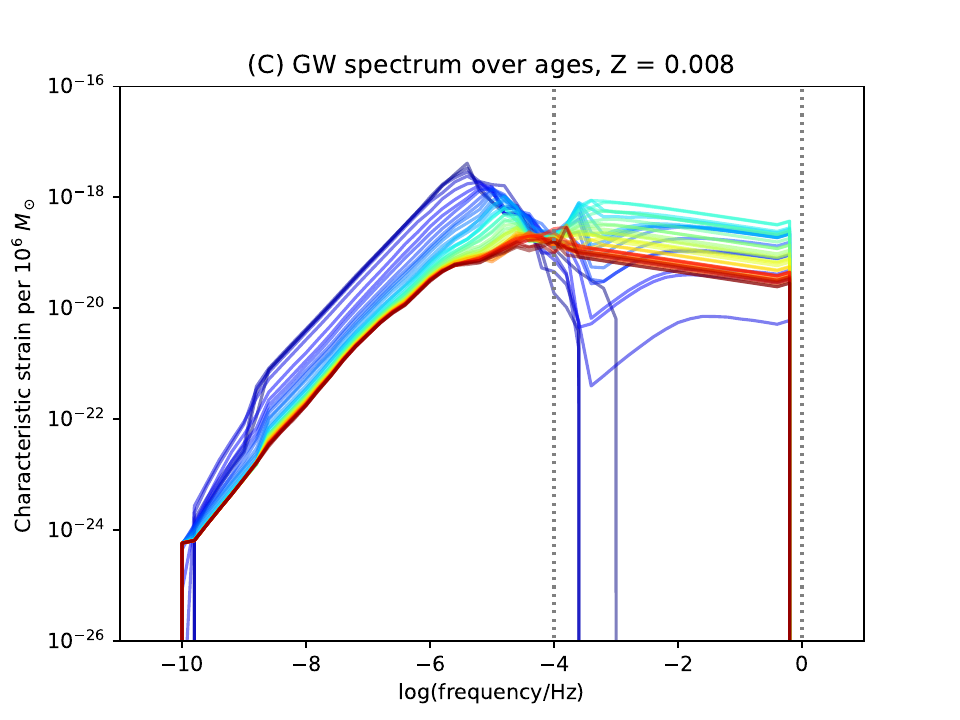}
    \includegraphics[width=1\columnwidth]{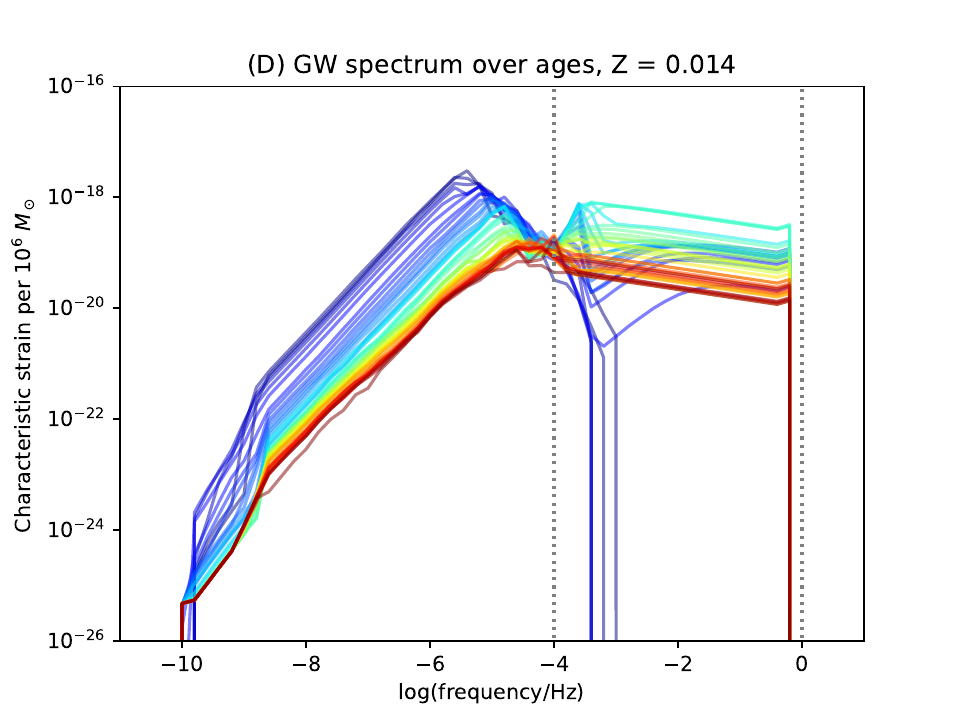}
    \raisebox{-0.5\height}{\includegraphics[width=1\columnwidth]{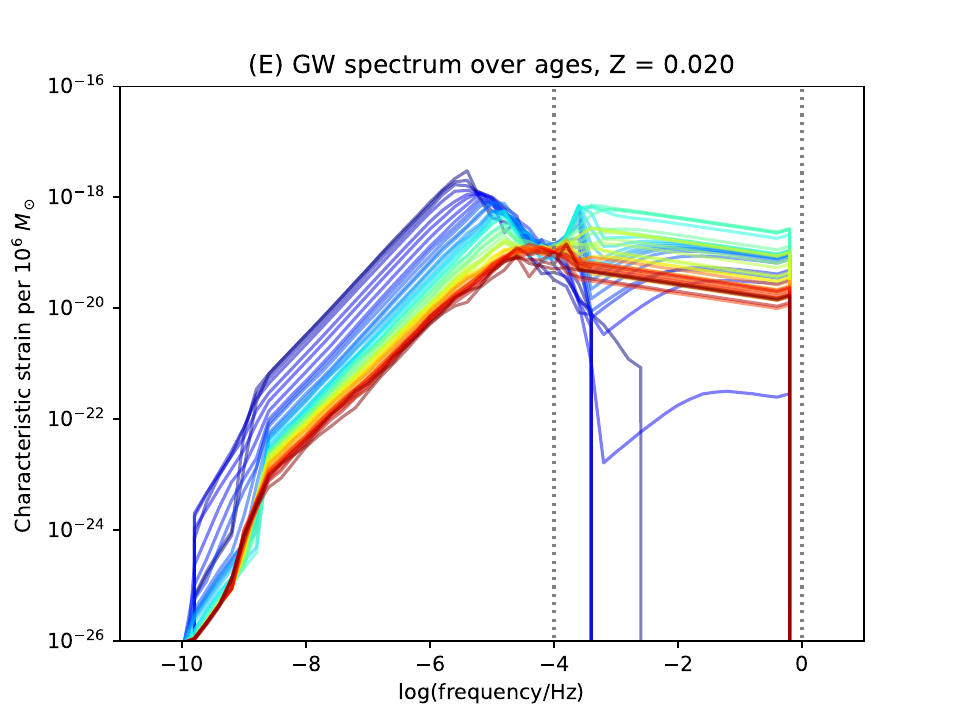}}
    \raisebox{-0.5\height}{\includegraphics[height=4.5cm]{colorbar.png}}
    \caption{Same as Figure \ref{evol_age}, but with characteristic strain plotted instead of strain PSD.}
    \label{evol_age_hc}
\end{figure*}

\section{Alternative arrangements of compact binary type plots}

Included in this appendix are alternative arrangements of the figures on different compact binary types, with each panel showing an individual age as opposed to an individual binary type.

\begin{figure*}
    \centering
    \includegraphics[width=1\columnwidth]{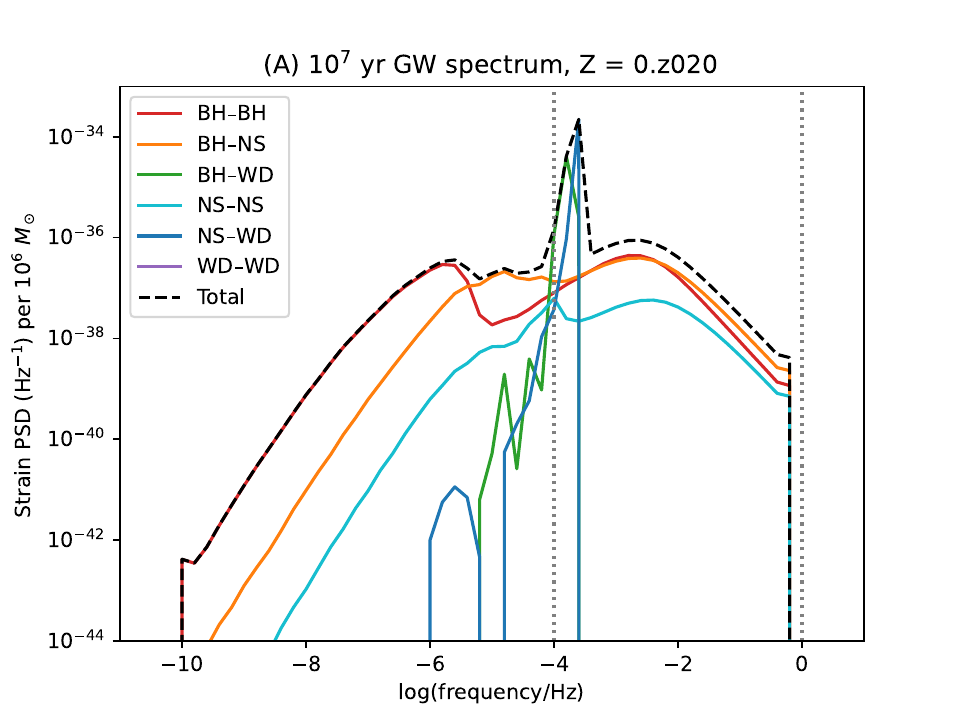}
    \includegraphics[width=1\columnwidth]{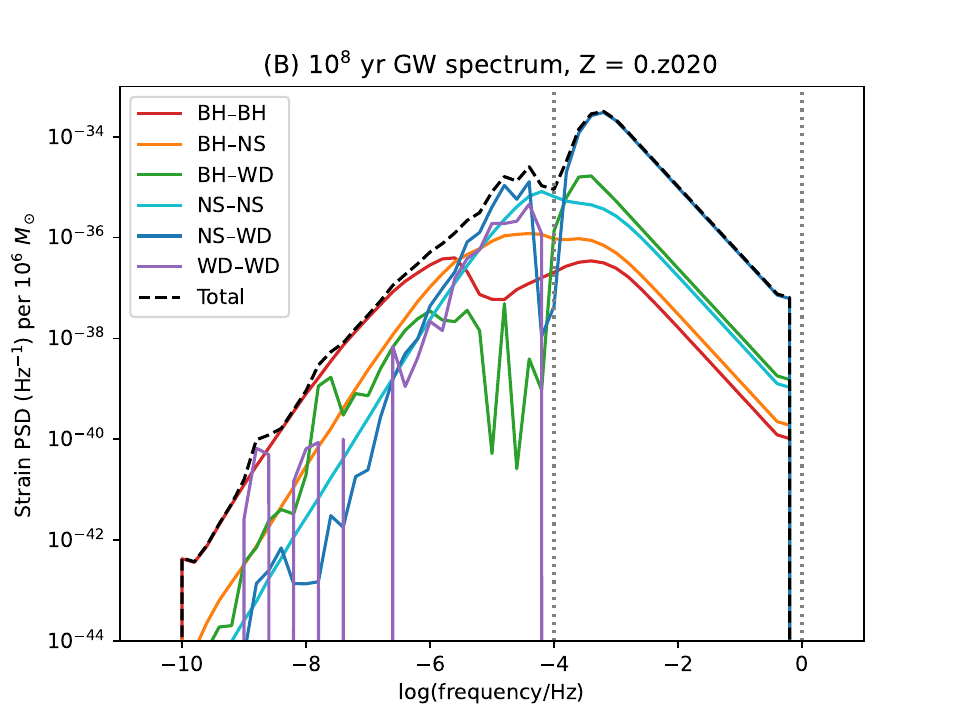}
    \includegraphics[width=1\columnwidth]{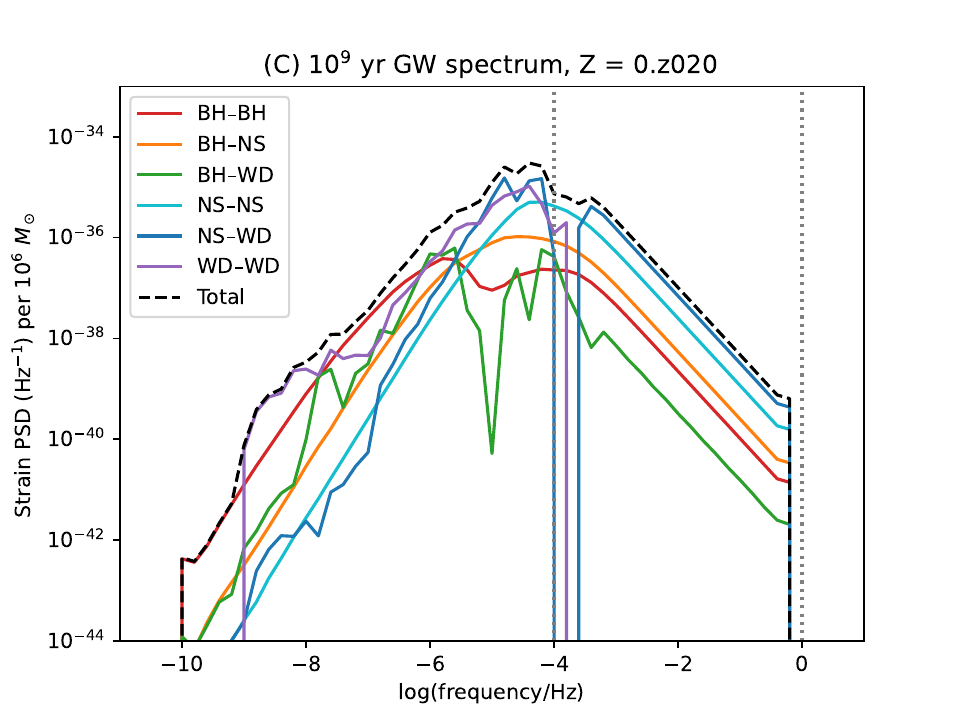}
    \includegraphics[width=1\columnwidth]{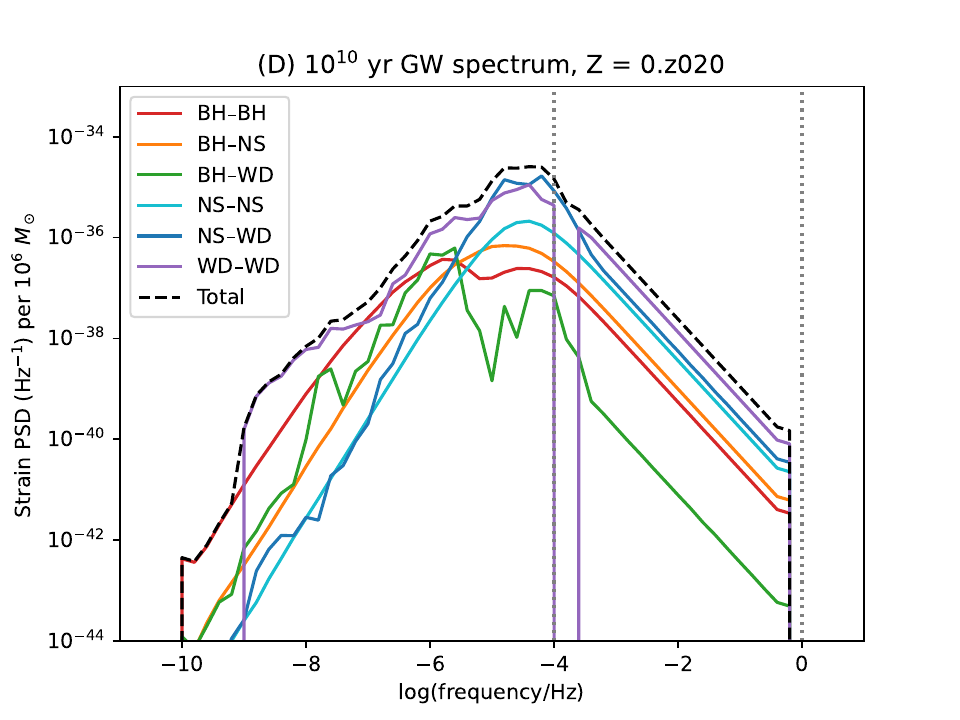}
    \includegraphics[width=1\columnwidth]{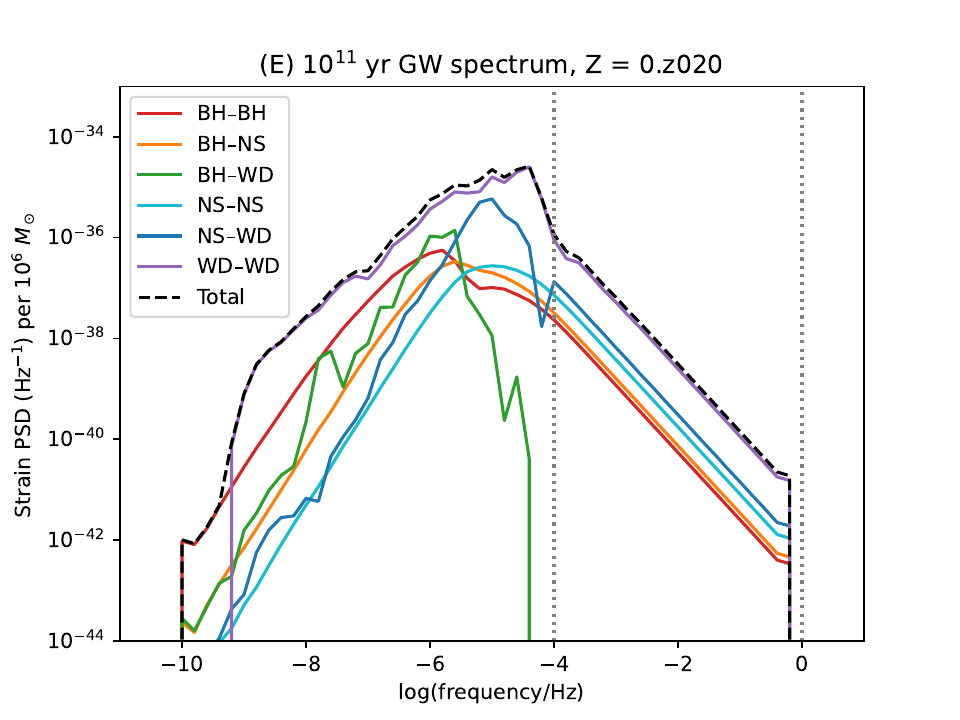}
    \caption{The same data as Figure \ref{compact_types}, but arranged with each panel showing a different point in time after the initial starburst, and each line a different binary type. The black dashed line shows the spectrum of all the compact remnant types added together. The dotted lines show an approximation of the LISA frequency range.}
    \label{compact_types_alt}
\end{figure*}

\begin{figure*}
    \centering
    \includegraphics[width=1\columnwidth]{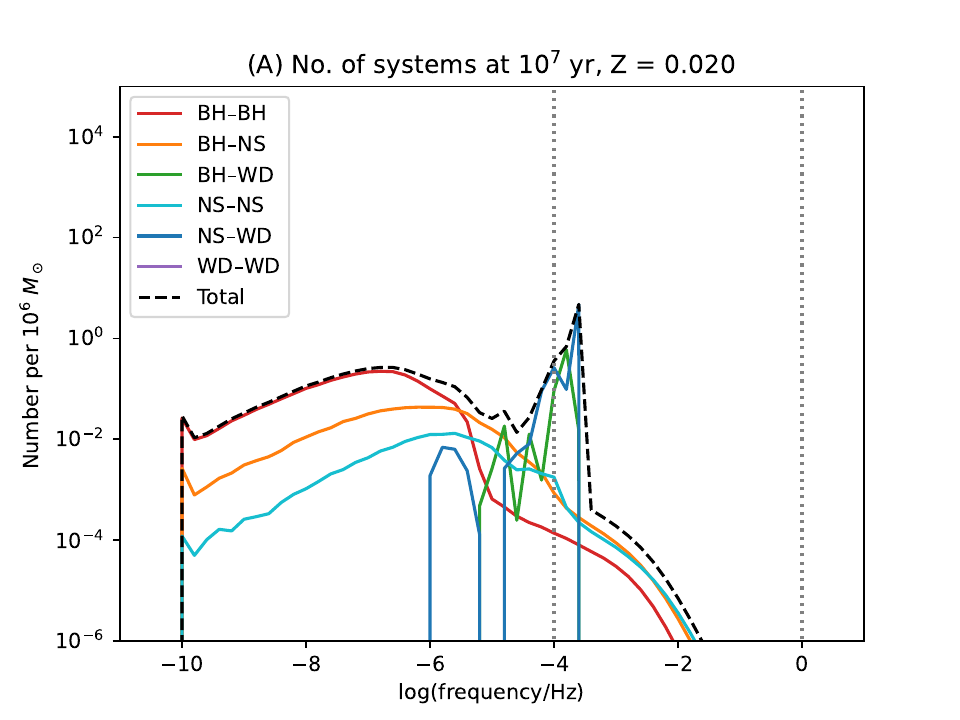}
    \includegraphics[width=1\columnwidth]{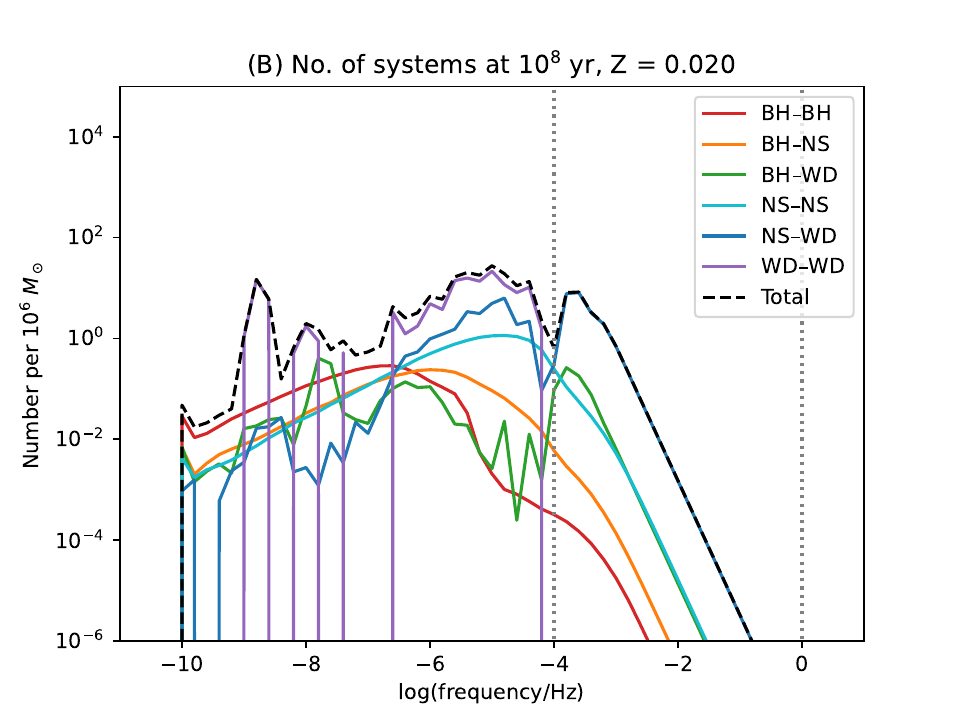}
    \includegraphics[width=1\columnwidth]{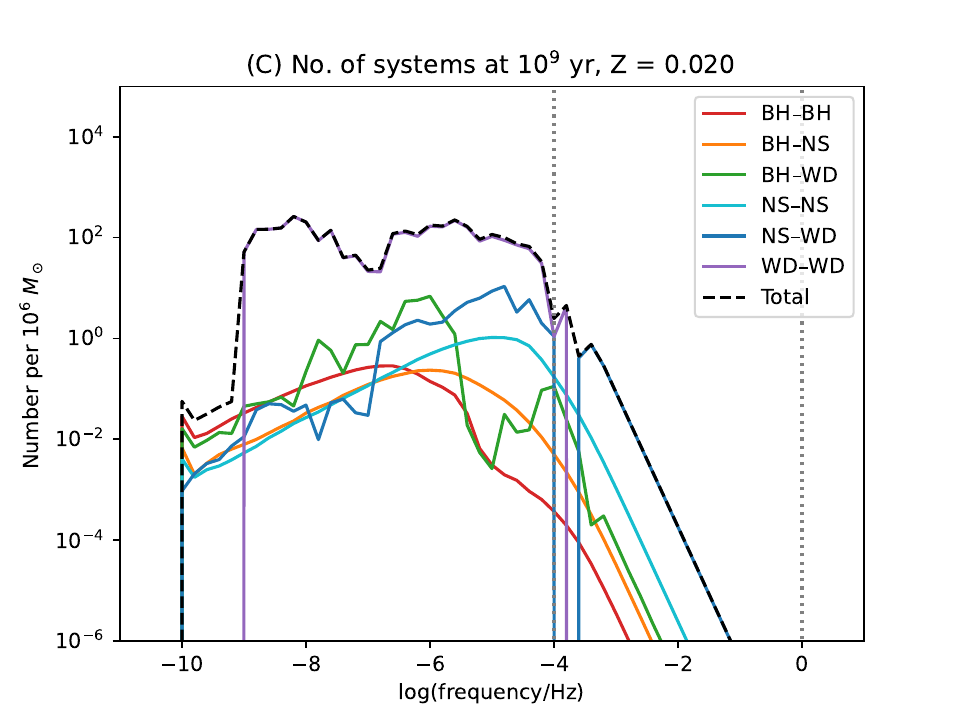}
    \includegraphics[width=1\columnwidth]{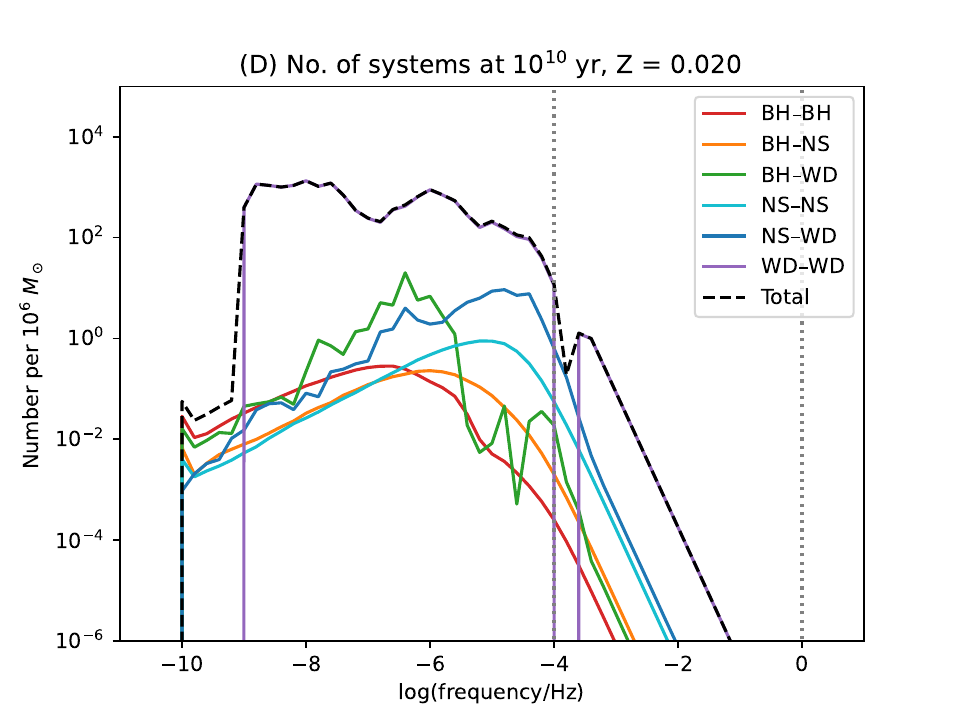}
    \includegraphics[width=1\columnwidth]{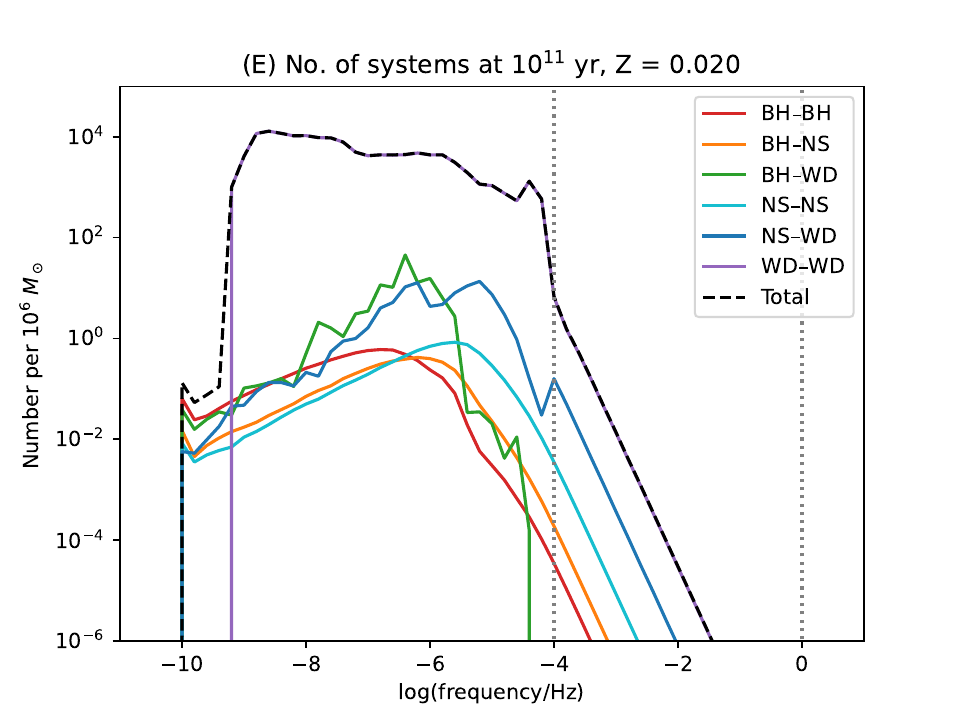}
    \caption{The same data as Figure \ref{system_no}, but arranged with each panel showing a different point in time after the initial starburst, and each line a different binary type. The black dashed line shows the numbers of all the compact remnant types added together. The dotted lines show an approximation of the LISA frequency range.}
    \label{system_no_alt}
\end{figure*}

\section{Data on open and globular clusters}

Included in this appendix are tables and a sky location plot showing the data on the real examples of open and globular clusters discussed in section \ref{chapter_examples}.

\begin{table*}
    \centering
    \begin{tabular}{ c | c c c c c c }
    Name & Ref. & Distance (pc) & log(Age / yr) & log(M$_{\rm current}$ / M$_\odot$) & Metallicity & Expected no. \\
    \hline
    Cep OB3 & a & 700 & 7.44 & 4.072 & 0.014* & 0.12 \\
    Melotte 20 & a & 190 & 7.55 & 4.186 & 0.014* & 0.31 \\
    Ruprecht 94 & a & 3400 & 7.19 & 4.463 & 0.014* & 0.13 \\
    VDB 113 & a & 3470 & 7.50 & 4.655 & 0.014* & 0.65 \\
    ASCC 88 & a & 1900 & 7.17 & 4.798 & 0.014* & 0.29 \\
    Sco OB5 & a & 3310 & 6.86 & 5.770 & 0.014* & 0.12 \\
    Nor OB5 & a & 1800 & 7.11 & 5.778 & 0.014* & 1.22 \\
    Trumpler 29 & b & 1350 & 7.728 & 3.710 & 0.020 & 0.12 \\
    Haffner 26 & b & 2520 & 8.769 & 4.163 & 0.010 & 0.13 \\
    
    \end{tabular}
    \caption{Data of open clusters from (a) \citet{catalogue_piskunov} and (b) \citet{catalogue_cordoni}, showing those that have an expected number of binaries in the LISA frequency range of at least 0.1.}
    \label{open_cluster_table}
\end{table*}

\begin{table*}
    \centering
    \begin{tabular}{ c | c c c c }
    Name & Distance (kpc) & log(Age / yr) & log(M$_{\rm current}$ / M$_\odot$) & Expected no. \\
    \hline
    $\omega$ Cen & 5.43 & 10.08 & 6.56 & 12.09 \\
    Terzan 5 & 6.62 & 10.08 & 5.97 & 3.11 \\
    Liller 1 & 8.06 & 10.08 & 5.96 & 3.04 \\
    47 Tuc & 4.52 & 10.07 & 5.95 & 2.97 \\
    M19 & 8.34 & 10.10 & 5.84 & 2.32 \\
    M62 & 6.41 & 10.06 & 5.79 & 2.03 \\
    M14 & 9.14 & 10.06 & 5.77 & 1.97 \\
    M13 & 7.42 & 10.08 & 5.74 & 1.81 \\
    NGC 6440 & 8.25 & 10.05 & 5.69 & 1.38 \\
    M22 & 3.30 & 10.10 & 5.68 & 1.58 \\
    M5 & 7.48 & 10.06 & 5.60 & 1.31 \\
    M92 & 8.50 & 10.10 & 5.55 & 1.17 \\
    NGC 6380 & 9.61 & 10.08 & 5.52 & 1.11 \\
    M9 & 8.30 & 10.08 & 5.51 & 1.07 \\
    M28 & 5.37 & 10.12 & 5.48 & 0.99 \\
    NGC 6541 & 7.61 & 10.10 & 5.47 & 0.97 \\
    NGC 6553 & 5.33 & 10.11 & 5.45 & 0.95 \\
    NGC 362 & 8.33 & 10.07 & 5.45 & 0.94 \\
    NGC 6752 & 4.12 & 10.10 & 5.44 & 0.92 \\
    NGC 5927 & 8.27 & 10.03 & 5.44 & 0.78 \\
    
    \end{tabular}
    \caption{Data of globular clusters from \citet{catalogue_baumgardt1} and \citet{catalogue_baumgardt2} and references therein, showing the 20 most massive clusters located within 10 kpc of Earth. Data retrieved from \url{https://people.smp.uq.edu.au/HolgerBaumgardt/globular/} on February 2023.}
    \label{globular_cluster_table}
\end{table*}

\begin{figure*}
    \centering
    \includegraphics[width=1.2\columnwidth]{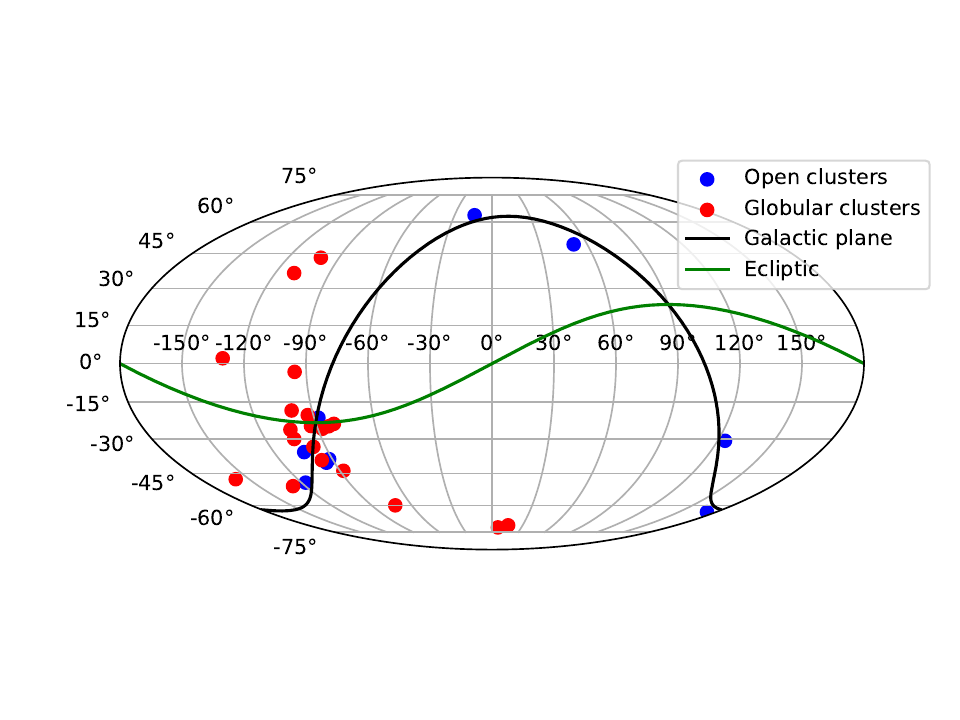}
    \caption{Sky locations of the open clusters in Table \ref{open_cluster_table} and the globular clusters in Table \ref{globular_cluster_table}, along with the Galactic and ecliptic planes.}
    \label{sky_location}
\end{figure*}


\bsp	
\label{lastpage}
\end{document}